\documentclass[11pt,twoside]{article}

\usepackage{fullpage}

\usepackage{amsfonts,graphicx}
\usepackage{amsmath,amssymb,amsthm}
\usepackage{fullpage}
\usepackage{epsfig}
\usepackage{epstopdf}
\usepackage{fancyheadings}
\usepackage{graphics}
\usepackage{amsfonts}
\usepackage{amsmath}
\usepackage{psfrag} \usepackage{enumerate} \usepackage{setspace}

\usepackage{color}

\theoremstyle{plain}

\newtheorem{theo}{Theorem}[section]

\newtheorem{lem}{Lemma}[section]
\newtheorem{prop}{Proposition}[section]
\newtheorem{cor}{Corollary}[section]

\theoremstyle{definition} 

\newtheorem{nota}{Notation}[section]
\newtheorem{de}{Definition}[section]
\newtheorem{exa}{Example}[section]
\newtheorem{as}{Assumption}[section]
\newtheorem{alg}{Algorithm}[section]

\newcommand{\btheo}{\begin{theo}}
\newcommand{\bde}{\begin{de}}
\newcommand{\ble}{\begin{lem}}
\newcommand{\bpr}{\begin{prop}}
\newcommand{\bno}{\begin{nota}}
\newcommand{\bex}{\begin{exa}}
\newcommand{\bcor}{\begin{cor}}
\newcommand{\spro}{\begin{proof}}
\newcommand{\bas}{\begin{as}}
\newcommand{\balg}{\begin{alg}}

\newcommand{\etheo}{\end{theo}}
\newcommand{\ede}{\end{de}}
\newcommand{\ele}{\end{lem}}
\newcommand{\epr}{\end{prop}}
\newcommand{\eno}{\end{nota}}
\newcommand{\eex}{\end{exa}}
\newcommand{\ecor}{\end{cor}}
\newcommand{\fpro}{\end{proof}}
\newcommand{\eas}{\end{as}}
\newcommand{\ealg}{\end{alg}}

\theoremstyle{plain}

\newtheorem{theos}{Theorem}
\newtheorem{props}{Proposition}
\newtheorem{lems}{Lemma}
\newtheorem{cors}{Corollary}

\theoremstyle{definition}
\newtheorem{exas}{Example}
\newtheorem{algs}{Algorithm}
\newtheorem{asss}{Assumption}
\newtheorem{defns}{Definition}

\newcommand{\btheos}{\begin{theos}}
\newcommand{\etheos}{\end{theos}}
\newcommand{\bprops}{\begin{props}}
\newcommand{\eprops}{\end{props}}
\newcommand{\bdes}{\begin{defns}}
\newcommand{\edes}{\end{defns}}
\newcommand{\blems}{\begin{lems}}
\newcommand{\elems}{\end{lems}}
\newcommand{\bcors}{\begin{cors}}
\newcommand{\ecors}{\end{cors}}
\newcommand{\bexs}{\begin{exas}}
\newcommand{\eexs}{\end{exas}}
\newcommand{\balgs}{\begin{algs}}
\newcommand{\ealgs}{\end{algs}}
\newcommand{\bass}{\begin{asss}}
\newcommand{\eass}{\end{asss}}

\newcommand{\numobs}{\ensuremath{n}}
\newcommand{\usedim}{\ensuremath{d}}
\newcommand{\kdim}{\ensuremath{k}}

\newcommand{\Sbar}{\ensuremath{S^c}}
\newcommand{\mprob}{\ensuremath{\mathbb{P}}}
\newcommand{\xstar}{\ensuremath{x^*}}
\newcommand{\xhat}{\ensuremath{\widehat{x}}}
\newcommand{\ehat}{\ensuremath{\widehat{e}}}
\newcommand{\numproj}{\ensuremath{m}}
\newcommand{\ConeSet}{\ensuremath{\mathcal{K}}}
\newcommand{\Sketch}{\ensuremath{S}}
\newcommand{\Width}{\ensuremath{\mathbb{W}}}
\newcommand{\WidthRad}{\ensuremath{\mathbb{R}}}
\newcommand{\Constraint}{\ensuremath{\mathcal{C}}}
\newcommand{\Amat}{\ensuremath{A}} \newcommand{\yvec}{\ensuremath{y}}
\newcommand{\xvec}{\ensuremath{x}}
\newcommand{\real}{\ensuremath{\mathbb{R}}}
\newcommand{\defn}{\ensuremath{: \, =}}

\newcommand{\inprod}[2]{\ensuremath{\langle #1 , \, #2 \rangle}}

\newcommand{\Sphere}[1]{\SPHERE{#1}}

\newcommand{\Exs}{\ensuremath{\mathbb{E}}}

\newcommand{\order}{\ensuremath{\mathcal{O}}}

\newcommand{\Sset}{\ensuremath{S}}
\newcommand{\sign}{\ensuremath{\mbox{sign}}}

\newcommand{\ConvSet}{\ensuremath{\mathcal{C}}}

\newcommand{\clconv}{\ensuremath{\operatorname{clconv}}}
\long\def\comment#1{}

\newcommand{\CEXP}[1]{\ensuremath{e^{#1}}}

\newcommand{\SPHERE}[1]{\ensuremath{\mathcal{S}^{#1-1}}}

\newcommand{\UNICON}{\ensuremath{c}}

\newcommand{\HACKPROB}{\ensuremath{1 - \UNICON_1 \CEXP{- \UNICON_2
      \numproj \delta^2}}}

\newcommand{\matsnorm}[2]{|\!|\!| #1 | \! | \!|_{{#2}}}

\newcommand{\nucnorm}[1]{\ensuremath{\matsnorm{#1}{\tiny{\mbox{nuc}}}}}

\newcommand{\frobnorm}[1]{\ensuremath{\matsnorm{#1}{\tiny{\mbox{fro}}}}}
\newcommand{\opnorm}[1]{\ensuremath{\matsnorm{#1}{\tiny{\mbox{op}}}}}

\newcommand{\HACKPROOF}{\begin{proof}}
\newcommand{\HACKENDPROOF}{\end{proof}}

\newcommand{\infset}{\ensuremath{\Amat \ConeSet \cap
    \SPHERE{\numobs}}}

\newcommand{\SUPERWIDTH}{\ensuremath{\Width(\Amat \ConeSet)}}

\newcommand{\SKETCHWIDTH}{\ensuremath{\Width_\Sketch(\Amat \ConeSet)}}
\newcommand{\PLSKETCHWIDTH}{\ensuremath{\Width_\Sketch}}
\newcommand{\SKETCHWIDTHSQ}{\ensuremath{\Width^2_\Sketch(\Amat
    \ConeSet)}}

\newcommand{\rank}{\operatorname{rank}}

\newcommand{\zhat}{\ensuremath{\widehat{z}}}

\newcommand{\MYEPS}{\ensuremath{\delta}}

\newcommand{\fixvec}{\ensuremath{u}}

\newcommand{\rade}[1]{\ensuremath{\varepsilon_{#1}}}
\newcommand{\sketch}{\ensuremath{s}}

\newcommand{\var}{\ensuremath{\operatorname{var}}}

\newcommand{\ZINF}{\ensuremath{Z_1(\Amat \ConeSet)}}
\newcommand{\ZSUP}{\ensuremath{Z_2(\Amat \ConeSet)}}
\newcommand{\ZSUPSTAR}{\ZSUP}

\newcommand{\MYLEMEPS}{\ensuremath{\MYEPS}}
\newcommand{\YSET}{\ensuremath{\mathcal{Y}}}
\newcommand{\VSET}{\ensuremath{\mathcal{V}}}

\newcommand{\cov}{\ensuremath{\operatorname{cov}}}

\newcommand{\Ball}{\ensuremath{\mathbb{B}}}

\newcommand{\Event}{\ensuremath{\mathcal{E}}}

\newcommand{\tracer}[2]{\ensuremath{\langle \!\langle {#1}, \; {#2}
    \rangle \!\rangle}} \newcommand{\rad}{\ensuremath{r}}

\newcommand{\matdim}{\ensuremath{d}}

 \newcommand{\DD}{\ensuremath{D}}
\newcommand{\PLGOOD}{\ensuremath{\mathcal{G}}}
\newcommand{\Ind}{\ensuremath{\mathbb{I}}}

\newcommand{\ZZERO}{\ensuremath{Z_0}}

\newcommand{\PLGOODONE}{\ensuremath{\PLGOOD_1}}
\newcommand{\PLGOODTWO}{\ensuremath{\PLGOOD_2}}

\newcommand{\diag}{\ensuremath{\mbox{diag}}}

\newcommand{\trunlev}{\ensuremath{\tau}}

\newcommand{\MYRAD}{\ensuremath{\mathbb{R}}}

\newcommand{\sketchtil}{\ensuremath{\widetilde{\sketch}}}

\newcommand{\trace}{\ensuremath{\mbox{trace}}}

\newcommand{\HACKKAP}{\ensuremath{\kappa}}

\newcommand{\STRANGEPROB}{\ensuremath{1 - \frac{c_1}{(\numproj
      \numobs)^2} - c_1 \exp \big(- c_2 \frac{\numproj
      \delta^2}{\WidthRad^2(\Amat \ConeSet) + \log(\numproj \numobs)}
    \big)}}

\newcommand{\STRANGEPROBY}{\ensuremath{1 - \frac{c_1}{(\numproj
      \numobs)^2} - c_1 \exp \big(- c_2 \frac{\numproj
      \delta^2}{\WidthRad^2(\YSET) + \log(\numproj \numobs)}
    \big)}}

\newcommand{\VSETPLUS}{\ensuremath{\VSET_+}}
\newcommand{\VSETMINUS}{\ensuremath{\VSET_-}}

\newcommand{\USET}{\ensuremath{\mathcal{U}}}
\newcommand{\USETPLUS}{\ensuremath{\USET_+}}
\newcommand{\USETMINUS}{\ensuremath{\USET_-}}

\newcommand{\SetDiff}[1]{\ensuremath{\partial[#1]}}
\newcommand{\SetDiffTwo}[1]{\ensuremath{\partial^2[#1]}}
\newcommand{\SphereProj}[1]{\ensuremath{\Pi(#1)}}

\newcommand{\conv}{\ensuremath{\operatorname{conv}}}

\newcommand{\SYMMAT}[1]{\ensuremath{\mathcal{S}^{#1 \times #1}}}
\newcommand{\radfloor}{\ensuremath{\lfloor \rad \rfloor}}

\newcommand{\sigkminsq}{\ensuremath{\gamma^-_{\kdim}}}
\newcommand{\siggroupminsq}{\ensuremath{\gamma^-_{\kdim, \GroupSet}}}
\newcommand{\sigkmaxsq}{\ensuremath{\gamma^+_{\kdim}}}
\newcommand{\sigrminsq}{\ensuremath{\gamma^-_{\rdim}}}
\newcommand{\sigrmaxsq}{\ensuremath{\gamma^+_{\rdim}}}

\newcommand{\Aop}{\ensuremath{\mathcal{A}}}
\newcommand{\myvec}{\ensuremath{\operatorname{vec}}}
\newcommand{\Xstar}{\ensuremath{X^*}}
\newcommand{\Xhat}{\ensuremath{\widehat{X}}}

\newcommand{\rdim}{\ensuremath{r}}

\newcommand{\ommin}{\ensuremath{\omega_{\tiny{\operatorname{min}}}}}
\newcommand{\ommax}{\ensuremath{\omega_{\tiny{\operatorname{max}}}}}
\newcommand{\usedima}{\ensuremath{\usedim_1}}
\newcommand{\usedimb}{\ensuremath{\usedim_2}}

\newcommand{\ROSHACK}{\ensuremath{1-e^{-c_1 \frac{\numproj
        \delta^2}{\log^4 \numobs}}}}

\newcommand{\ROSHACKN}{\ensuremath{1-e^{-c_1 \frac{\numproj
        \delta^2}{\log^4 \numobs}}}}

\newcommand{\ROSHACKMAT}{\ensuremath{1-e^{-c_1 \frac{\numproj
        \delta^2}{\log^4 (\usedima \, \usedimb)}}}}

\newcommand{\widgraph}[2]{\includegraphics[keepaspectratio,width=#1]{#2}}

\newcommand{\MYTEXTWIDTH}{.8\textwidth}
\newcommand{\uvec}{\ensuremath{u}}

\newcommand{\GroupSet}{\ensuremath{\mathcal{G}}}
\newcommand{\group}{\ensuremath{g}}
\newcommand{\gmax}{\ensuremath{M}}
\makeatletter
\long\def\@makecaption#1#2{
        \vskip 0.8ex
        \setbox\@tempboxa\hbox{\small {\bf #1:} #2}
        \parindent 1.5em  
        \dimen0=\hsize
        \advance\dimen0 by -3em
        \ifdim \wd\@tempboxa >\dimen0
                \hbox to \hsize{
                        \parindent 0em
                        \hfil 
                        \parbox{\dimen0}{\def\baselinestretch{0.96}\small
                                {\bf #1.} #2
                                } 
                        \hfil}
        \else \hbox to \hsize{\hfil \box\@tempboxa \hfil}
        \fi
        }
\makeatother

\begin{document}

\begin{center}
  {\LARGE{\bf{ Randomized Sketches of Convex Programs \\ with Sharp
        Guarantees}}}

  \vspace{1cm}

  {\large
\begin{tabular}{ccc}
Mert Pilanci$^\star$ & Martin J. Wainwright$^{\star, \dagger}$
\end{tabular}
}

  \vspace{.5cm}

  \texttt{\{mert,wainwrig\}@berkeley.edu} \\

  \vspace{.5cm}

  {\large $^\star$Department of Electrical Engineering and Computer
    Science ~~~~ $^\dagger$Department of Statistics} \\
\vspace{.1cm}

  {\large University of California, Berkeley} \\

  \vspace{.5cm}

\today
\end{center}

\vspace*{.2cm}

\begin{abstract}
Random projection (RP) is a classical technique for reducing storage
and computational costs.  We analyze RP-based approximations of convex
programs, in which the original optimization problem is approximated
by the solution of a lower-dimensional problem.  Such dimensionality
reduction is essential in computation-limited settings, since the
complexity of general convex programming can be quite high (e.g.,
cubic for quadratic programs, and substantially higher for
semidefinite programs). In addition to computational savings, random
projection is also useful for reducing memory usage, and has useful
properties for privacy-sensitive optimization.  We prove that the
approximation ratio of this procedure can be bounded in terms of the
geometry of constraint set.  For a broad class of random projections,
including those based on various sub-Gaussian distributions as well as
randomized Hadamard and Fourier transforms, the data matrix defining
the cost function can be projected down to the statistical dimension
of the tangent cone of the constraints at the original solution, which
is often substantially smaller than the original dimension.  We
illustrate consequences of our theory for various cases, including
unconstrained and $\ell_1$-constrained least squares, support vector
machines, low-rank matrix estimation, and discuss implications on
privacy-sensitive optimization and some connections with denoising and
compressed sensing.
\end{abstract}


\section{Introduction}

Optimizing a convex function subject to constraints is fundamental to
many disciplines in engineering, applied mathematics, and
statistics~\cite{Boyd02,Nesterov04}.  While most convex programs can
be solved in polynomial time, the computational cost can still be
prohibitive when the problem dimension and/or number of constraints
are large.  For instance, although many quadratic programs can be
solved in cubic time, this scaling may be prohibitive when the
dimension is on the order of millions.  This type of concern is only
exacerbated for more sophisticated cone programs, such as second-order
cone and semidefinite programs.  Consequently, it is of great interest
to develop methods for approximately solving such programs, along with
rigorous bounds on the quality of the resulting approximation.

In this paper, we analyze a particular scheme for approximating a
convex program defined by minimizing a quadratic objective function
over an arbitrary convex set.  The scheme is simple to describe and
implement, as it is based on performing a random projection of the
matrices and vectors defining the objective function.  Since the
underlying constraint set may be arbitrary, our analysis encompasses
many problem classes including quadratic programs (with constrained or
penalized least-squares as a particular case), as well as second-order
cone programs and semidefinite programs (including low-rank matrix
approximation as a particular case).

An interesting class of such optimization problems arise in the
context of statistical estimation.  Many such problems can be
formulated as estimating an unknown parameter based on noisy linear
measurements, along with side information that the true parameter
belongs to a low-dimensional space.  Examples of such low-dimensional
structure include sparse vectors, low-rank matrices, discrete sets
defined in a combinatorial manner, as well as algebraic sets,
including norms for inducing shrinkage or smoothness.  Convex
relaxations provide a principled way of deriving polynomial-time
methods for such problems~\cite{Boyd02}, and their statistical
performance has been extensively studied over the past decade (see the
papers~\cite{ChaRec12,Wai14} for overviews).  For many such problems,
the ambient dimension of the parameter is very large, and the number
of samples can also be large.  In these contexts, convex programs may
be difficult to solve exactly, and reducing the dimension and sample
size by sketching is a very attractive option.

Our work is related to a line of work on sketching unconstrained
least-squares problems (e.g., see the
papers~\cite{DriMahMutSar09,Mahoney11,Boutsidis13} and references
therein).  The results given here generalizes this line of work by
providing guarantees for the broader class of constrained quadratic
programs.  In addition, our techniques are convex-analytic in nature,
and by exploiting analytical tools from Banach space geometry and
empirical process theory~\cite{PenGin, LedTal91, Ledoux01}, lead to
sharper bounds on the sketch size as well as sharper probabilistic
guarantees.  Our work also provides a unified view of both
least-squares sketching~\cite{DriMahMutSar09,Mahoney11,Boutsidis13}
and compressed sensing~\cite{Donoho06,Donoho13}.  As we discuss in the
sequel, various results in compressed sensing can be understood as
special cases of sketched least-squares, in which the data matrix in
the original quadratic program is the identity.

In addition to reducing computation and storage, random projection is
also useful in the context of privacy preservation.  Many types of
modern data, including financial records and medical tests, have
associated privacy concerns.  Random projection allows for a sketched
version of the data set to be stored, but such that there is a
vanishingly small amount of information about any given data point.
Our theory shows that this is still possible, while still solving a
convex program defined by the data set up to $\delta$-accuracy.  In
this way, we sharpen some results by Zhou and Wasserman~\cite{Zhou09}
on privacy-preserving random projections for sparse regression.  Our
theory points to an interesting dichotomy in privacy sensitive
optimization problems based on the trade-off between the complexity of
the constraint set and mutual information. We show that if the
constraint set is \emph{simple} enough in terms of a statistical
measure, privacy sensitive optimization can be done with arbitrary
accuracy.

The remainder of this paper is organized as follows.  We begin in
Section~\ref{SecMain} with a more precise formulation of the problem,
and the statement of our main results.  In Section~\ref{SecConcrete},
we derive corollaries for a number of concrete classes of problems,
and provide various simulations that demonstrate the close agreement
between the theoretical predictions and behavior in practice.
Sections~\ref{SecProofs} and Section~\ref{SecSharpen} are devoted to
the proofs our main results, and we conclude in
Section~\ref{SecDiscussion}.  Parts of the results given here are to
appear in the conference form at the International Symposium on
Information Theory (2014).



\section{Statement of main results}
\label{SecMain}

We begin by formulating the problem analyzed in this paper, before
turning to a statement of our main results.


\subsection{Problem formulation}

Consider a convex program of the form
\begin{align}
\label{EqnOriginalProblem}
\xstar & \in \arg \min_{x \in \Constraint} \underbrace{\|\Amat \xvec -
  \yvec\|_2^2}_{f(\xvec)},
\end{align}
where $\Constraint$ is some convex subset of $\real^\usedim$, and
$\yvec \in \real^\numobs$ $\Amat \in \real^{\numobs \times \usedim}$
are a data vector and data matrix, respectively. Our goal is to obtain
an $\MYEPS$-optimal solution to this problem in a computationally
simpler manner, and we do so by projecting the problem into
$\real^\numproj$, where $\numproj < \numobs$, via a \emph{sketching
  matrix} $\Sketch \in \real^{\numproj \times \numobs}$.  In
particular, consider the \emph{sketched problem}
\begin{align}
\label{EqnSketchedProblem}
\xhat & \in \arg \min_{\xvec \in \Constraint} \underbrace{\|\Sketch
  (\Amat \xvec - \yvec)\|_2^2}_{g(\xvec)}.
\end{align}
Note that by the optimality and feasibility of $\xstar$ and $\xhat$,
respectively, for the original problem~\eqref{EqnOriginalProblem}, we
always have $f(\xstar) \leq f(\xhat)$.  Accordingly, we say that
$\xhat$ is an \emph{$\MYEPS$-optimal approximation} to the original
problem~\eqref{EqnOriginalProblem} if
\begin{align}
\label{EqnDefnEpsOptimal}
f(\xhat) & \leq \big (1 + \MYEPS \big)^2 \, f(\xstar).
\end{align}
Our main result characterizes the number of samples $\numproj$
required to achieve this bound as a function of $\MYEPS$, and other
problem parameters.

Our analysis involves a natural geometric object in convex analysis,
namely the tangent cone of the constraint set $\ConvSet$ at the
optimum $\xstar$, given by
\begin{align}
\label{EqnDefnTangentCone}
\ConeSet \defn & \clconv \big \{ \Delta \in \real^\usedim \, \mid \,
\mbox{$\Delta = t (x - \xstar)$ for some $t \geq 0$ and $x \in
  \ConvSet$} \},
\end{align}
where $\clconv$ denotes the closed convex hull.  This set arises
naturally in the convex optimality conditions for the original
problem~\eqref{EqnOriginalProblem}: any vector $\Delta \in \ConeSet$
defines a feasible direction at the optimal $\xstar$, and optimality
means that it is impossible to decrease the cost function by moving in
directions belonging to the tangent cone.

We use $\Amat \ConeSet$ to denote the linearly transformed cone $\{
\Amat \Delta \in \real^\numobs \, \mid \, \Delta \in \ConeSet\}$.  Our
main results involve measures of the ``size'' of this transformed cone
when it is intersected with the Euclidean sphere $\SPHERE{\numobs} =
\{ z \in \real^\numobs \, \mid \, \|z\|_2 = 1 \}$.  In particular, we
define \emph{Gaussian width} of the set \mbox{$\Amat \ConeSet \cap
  \SPHERE{\numobs}$} via
\begin{align}
\label{EqnDefnGaussWidth}
\Width(\Amat \ConeSet) & \defn \Exs_g \big[ \sup_{z \in \infset} \big|
  \inprod{g}{z} \big| \big]
\end{align}
where $g \in \real^\numobs$ is an i.i.d. sequence of $N(0,1)$
variables.  This complexity measure plays an important role in Banach
space theory, learning theory and statistics
(e.g.,~\cite{Pisier86,LedTal91,Bar05}).

\subsection{Guarantees for sub-Gaussian sketches}

Our first main result provides a relation between the sufficient
sketch size and Gaussian complexity in the case of sub-Gaussian
sketches.  In particular, we say that a row $\sketch_i$ of the
sketching matrix is \emph{$\sigma$-sub-Gaussian} if it is zero-mean,
and if for any fixed unit vector $u \in \SPHERE{\numobs}$, we have
\begin{align}
\label{EqnDefnSubgauss}
\mprob \big[ |\inprod{u}{\sketch_i}| \geq t] & \leq 2 \CEXP{-
  \frac{\numobs t^2}{2 \sigma^2}} \quad \mbox{for all $t \geq 0$.}
\end{align}
Of course, this condition is satisfied by the standard Gaussian sketch
($\sketch_i \sim N(0, I_{\numobs \times \numobs})$).  In addition, it
holds for various other sketching matrices, including random matrices
with i.i.d.  Bernoulli elements, random matrices with rows drawn
uniformly from the rescaled unit sphere, and so on.  We say that the
sketching matrix $\Sketch \in \real^{\numproj \times \numobs}$ is
drawn from a $\sigma$-sub-Gaussian ensemble if each row is
$\sigma$-sub-Gaussian in the previously defined
sense~\eqref{EqnDefnSubgauss}.
\btheos[Guarantees for sub-Gaussian projections]
\label{ThmMain}
Let $\Sketch \in \real^{\numproj \times \numobs}$ be drawn from a
$\sigma$-sub-Gaussian ensemble.  Then there are universal constants
$(\UNICON_0, \UNICON_1, \UNICON_2)$ such that, for any tolerance parameter
$\MYEPS \in (0,1)$, given a sketch size lower bounded as
\begin{align}
\label{EqnMainResult}
\numproj & \geq \frac{\UNICON_0}{\MYEPS^2} \, \Width^2(\Amat\ConeSet),
\end{align}
the approximate solution $\xhat$ is guaranteed to be
$\MYEPS$-optimal~\eqref{EqnDefnEpsOptimal} for the original program
with probability at least $\HACKPROB$.
\etheos

\noindent As will be clarified in examples to follow, the squared
width $\Width^2(\Amat \ConeSet)$ scales proportionally to the
statistical dimension, or number of degrees of freedom in the set
$\infset$.  Consequently, up to constant factors,
Theorem~\ref{ThmMain} guarantees that we can project down to the
statistical dimension of the problem while preserving
$\delta$-optimality of the solution. \\

This fact has an interesting corollary in the context of
privacy-sensitive optimization.  Suppose that we model the data matrix
$\Amat \in \real^{\numobs \times \usedim}$ as being random, and our
goal is to solve the original convex
program~\eqref{EqnOriginalProblem} up to $\delta$-accuracy while
revealing as little as possible about the individual entries of
$\Amat$.  By Theorem~\ref{ThmMain}, whenever the sketch dimension
satisfies the lower bound~\eqref{EqnMainResult}, the sketched data
matrix $\Sketch \Amat \in \real^{\numproj \times \usedim}$ suffices to
solve the original program up to $\delta$-accuracy.  We can thus ask
about how much information per entry of $\Amat$ is retained by the
sketched data matrix.  One way in which to do so is by computing the
mutual information per symbol, namely
\begin{align*}
\frac{I( \Sketch \Amat; \Amat)}{\numobs \usedim} & = \frac{1}{\numobs
  \usedim} \big \{ H(\Amat) - H( \Amat \, \mid \, \Sketch \Amat) \big
\},
\end{align*}
where the rescaling is chosen since $\Amat$ has a total of $\numobs
\usedim$ entries.  This quantity was studied by Zhou and
Wasserman~\cite{Zhou09} in the context of privacy-sensitive sparse
regression, in which $\Constraint$ is an $\ell_1$-ball, to be
discussed at more in length in Section~\ref{SecLasso}.  In our
setting, we have the following more generic corollary of Theorem~\ref{ThmMain}:
\bcors
\label{CorPrivate}
Let the entries of $\Amat$ be drawn i.i.d. from a distribution with
finite variance $\gamma^2$.  Byusing $\numproj =
\frac{\UNICON_0}{\MYEPS^2} \, \Width^2(\Amat \ConeSet)$ random
Gaussian projections, we can ensure that
\begin{align}
\label{EqnMutInfoBound}
\frac{I(\Sketch \Amat; \Amat)}{\numobs \usedim} & \leq
\frac{\UNICON_0}{\MYEPS^2} \, \frac{\Width^2(\Amat
  \ConeSet)}{\numobs} \log(2 \pi e \gamma^2),
\end{align}
and that the sketched solution is $\MYEPS$-optimal with probability
at least $\HACKPROB$.
\ecors
\noindent
Note that the inequality $\Width^2(\Amat \ConeSet) \leq \numobs$
always holds.  However, for many problems, we have the much stronger
guarantee $\Width^2(\Amat \ConeSet) = o(\numobs)$, in which case the
bound~\eqref{EqnMutInfoBound} guarantees that the mutual information
per symbol is vanishing.  There are various concrete problems, as
discussed in Section~\ref{SecConcrete}, for which this type of scaling
is reasonable.  Thus, for any fixed $\MYEPS \in (0,1)$, we are
guaranteed a $\MYEPS$-optimal solution with a vanishing mutual
information per symbol.

Corollary~\ref{CorPrivate} follows by a straightforward combination of
past work~\cite{Zhou09} with Theorem~\ref{ThmMain}.  Zhou and
Wasserman~\cite{Zhou09} show that under the stated conditions, for a
standard i.i.d. Gaussian sketching matrix $S$, the mutual information
rate per symbol is upper bounded as
\begin{align*}
\frac{I(\Sketch \Amat; \Amat)}{\numobs \usedim} & \leq
\frac{\numproj}{2\numobs} \, \log(2\pi e \gamma^2).
\end{align*}
Substituting in the stated choice of $\numproj$ and applying
Theorem~\ref{ThmMain} yields the claim.


\subsection{Guarantees for randomized orthogonal systems}

A possible disadvantage of using sub-Gaussian sketches is that it
requires performing matrix-vector multiplications with unstructured
random matrices; such multiplications require $\order(mnd)$ time in
general.  Our second main result applies to sketches based on a
\emph{randomized orthonormal system} (ROS), for which matrix
multiplication can be performed much more quickly.

In order to define a randomized orthonormal system, we begin by with
an orthonormal matrix $H \in \real^{\numobs \times \numobs}$ with
entries $H_{ij} \in \{ -\frac{1}{\sqrt{\numobs}},
\frac{1}{\sqrt{\numobs}} \}$.  A standard class of such matrices is
provided by the Hadamard basis, for which matrix-vector multiplication
can be performed in $\order(\numobs \log \numobs)$ time.  Another
possible choice is the Fourier basis.  Based on any such matrix, a
sketching matrix $\Sketch \in \real^{\numproj \times \numobs}$ from a
ROS ensemble is obtained by sampling i.i.d. rows of the form
\begin{align*}
\sketch_i = \sqrt{\numobs} D H^T p_i,
\end{align*}
where the random vector $p_i \in \real^\numobs$ is chosen uniformly at
random from the set of all $\numobs$ canonical basis vectors, and $D =
\diag(\nu)$ is a diagonal matrix of i.i.d. Rademacher variables $\nu
\in \{-1, +1\}^\numobs$.  With the base matrix $H$ chosen as the
Hadamard or Fourier basis, then for any fixed vector $x \in
\real^\numobs$, the product $\Sketch x$ can be computed in
$\order(\numobs \log \numproj)$ time (e.g., see the
paper~\cite{Ailon08} for details). Hence the sketched data $(\Sketch
A, \Sketch y)$ can be formed in $\order(\usedim \, \numobs \, \log
\numproj)$ time, which scales almost linearly in the input size
$\usedim \numobs$.

Our main result for randomized orthonormal systems involves the
\emph{$\Sketch$-Gaussian width} of the set \mbox{$\Amat \ConeSet \cap
  \SPHERE{\numobs}$}, given by
\begin{align}
\label{EqnDefnSketchWidth}
\SKETCHWIDTH & \defn \Exs_{g, \Sketch} \Big[ \sup_{z \in \infset}
  \Big| \inprod{g}{\frac{\Sketch z}{\sqrt{\numproj}}} \Big| \Big].
\end{align}
As will be clear in the corollaries to follow, in many cases, the
$\Sketch$-Gaussian width is equivalent to the ordinary Gaussian
width~\eqref{EqnDefnGaussWidth} up to numerical constants.  It also
involves the \emph{Rademacher width} of the set $\infset$, given by
\begin{align}
\WidthRad(\Amat \ConeSet) & = \Exs_{\rade{}} \big[ \sup_{z \in
    \infset} \big| \inprod{z}{\rade{}} \big| \big],
\end{align}
where $\rade{} \in \{-1, +1\}^\numobs$ is an i.i.d. vector of Rademacher
variables.
\btheos[Guarantees for randomized orthonormal system]
\label{ThmHadamard}
Let $\Sketch \in \real^{\numproj \times \numobs}$ be drawn from a
randomized orthonormal system (ROS).  Then given a sample size
$\numproj$ lower bounded as
\begin{align}
\label{EqnKeyLower}
\frac{\numproj}{\log \numproj} > \frac{\UNICON_0}{\delta^2} \big(
\WidthRad^2(\Amat \ConeSet) + \log \numobs \big) \; \SKETCHWIDTHSQ,
\end{align}
the approximate solution $\xhat$ is guaranteed to be
$\MYEPS$-optimal~\eqref{EqnDefnEpsOptimal} for the original program
with probability \hfill \mbox{at least $\STRANGEPROB$.}
\etheos
The required projection dimension~\eqref{EqnKeyLower} for ROS sketches
is in general larger than that required for sub-Gaussian sketches, due
to the presence of the additional pre-factor $\WidthRad^2(\Amat
\ConeSet) + \log \numobs$.  For certain types of cones, we can use
more specialized techniques to remove this pre-factor, so that it is
not always required.  The details of these arguments are given in
Section~\ref{SecSharpen}, and we provide some illustrative examples of
such sharpened results in the corollaries to follow.  However, the
potentially larger projection dimension is offset by the much lower
computational complexity of forming matrix vector products using the
ROS sketching matrix.


\section{Some concrete instantiations}
\label{SecConcrete}

Our two main theorems are general results that apply to any choice of
the convex constraint set $\Constraint$.  We now turn to some
consequences of Theorems~\ref{ThmMain} and~\ref{ThmHadamard} for more
specific classes of problems, in which the geometry enters in
different ways.

\subsection{Unconstrained least squares}  

We begin with the simplest possible choice, namely $\Constraint =
\real^\usedim$, which leads to an unconstrained least squares problem.
This class of problems has been studied extensively in past work on
least-square sketching~\cite{Mahoney11}; our derivation here provides
a sharper result in a more direct manner.  At least intuitively, given
the data matrix $\Amat \in \real^{\numobs \times \usedim}$, it should
be possible to reduce the dimensionality to the rank of the data
matrix $\Amat$, while preserving the accuracy of the solution.  In
many cases, the quantity $\rank(\Amat)$ is substantially smaller than
$\max \{ \numobs, \usedim\}$.  The following corollaries of
Theorem~\ref{ThmMain} and~\ref{ThmHadamard} confirm this intuition:
\bcors[Approximation guarantee for unconstrained least squares] \label{CorUncLS}
\begin{enumerate}
Consider the case of unconstrained least squares with $\Constraint = \real^\usedim$:
\item[(a)] Given a sub-Gaussian sketch with dimension $\numproj >
  \UNICON_0 \, \frac{\rank(\Amat)}{\MYEPS^2}$, the sketched solution
  is $\MYEPS$-optimal~\eqref{EqnDefnEpsOptimal} with probability at
  least $\HACKPROB$.
\item[(b)] Given a ROS sketch with dimension $\numproj > \UNICON_0'
  \frac{\rank(\Amat)}{\MYEPS^2} \, \log^4(\numobs) $, the sketched
  solution is \mbox{$\MYEPS$-optimal~\eqref{EqnDefnEpsOptimal}} with
  probability at least $\HACKPROB$.
\end{enumerate}
\ecors
\noindent This corollary improves known results both in the
probability estimate and required samples, in particular previous
results hold only with constant probability; see the
paper~\cite{Mahoney11} for an overview of such results. Note that the
total computational complexity of computing $SA$ and solving the
sketched least squares problem, for instance via QR decomposition
\cite{Golub96}, is of the order $\order(\numobs\usedim\numproj +
\numproj \usedim^2)$ for sub-Gaussian sketches, and of the order
$\order(\numobs \usedim \log(\numproj)+\numproj \usedim^2)$ for ROS
sketches.  Consequently, by using ROS sketches, the overall complexity
of computing a $\delta$-approximate least squares solution with
exponentially high probability is $\order(\rank(A) \usedim^2
\log^4(\numobs)/\delta^2 + \numobs\usedim\log(\rank(A)/\delta^2))$.
In many cases, this complexity is substantially lower than direct
computation of the solution via QR decomposition, which would require
$\order(\numobs\usedim^2)$ operations.

\HACKPROOF
Since $\Constraint = \real^\usedim$, the tangent cone $\ConeSet$ is all
of $\real^\usedim$, and the set $\Amat \ConeSet$ is the image of
$\Amat$.  Thus, we have
\begin{align}
\label{EqnSubspaceWidth}
\SUPERWIDTH & = \Exs \big[ \sup_{u \in \real^\usedim} \frac{
    \inprod{\Amat u}{g}}{\|\Amat u\|_2} \big] \; \leq \;
\sqrt{\rank(\Amat)},
\end{align}
where the inequality follows from the the fact that the image of
$\Amat$ is at most $\rank(\Amat)$-dimensional.  Thus, the sub-Gaussian
bound in part (a) is an immediate consequence of Theorem~\ref{ThmMain}.

Turning to part (b), an application of Theorem~\ref{ThmHadamard} will
lead to a sub-optimal result involving $(\rank(A))^2$.  In
Section~\ref{SecSubspaceSharpen}, we show how a refined argument will
lead to bound stated here.
\HACKENDPROOF

\newcommand{\CONLS}{\ensuremath{1.5}}
\newcommand{\CONLAS}{\ensuremath{4}}
\newcommand{\CONSVM}{\ensuremath{5}}

In order to investigate the theoretical prediction of
Corollary~\ref{CorUncLS}, we performed some simple simulations on randomly generated problem instances.
Fixing a dimension $\usedim = 500$, we formed a random ensemble of
least-squares problems by first generating a random data matrix $\Amat
\in \real^{n \times 500}$ with i.i.d. standard Gaussian entries.  For
a fixed random vector $x_0 \in \real^\usedim$, we then computed the
data vector $y = \Amat x_0 + w$, where the noise vector $w \sim N(0,
\nu^2)$ where $\nu = \sqrt{0.2}$.  Given this random ensemble of
problems, we computed the projected data matrix-vector pairs $(\Sketch
\Amat, \Sketch y)$ using Gaussian, Rademacher, and randomized Hadamard
sketching matrices, and then solved the projected convex program.  We
performed this experiment for a range of different problem sizes
$\numobs \in \{1024, 2048, 4096 \}$.  For any $\numobs$ in this set,
we have $\rank(\Amat) = \usedim = 500$, with high probability over the
choice of randomly sampled $\Amat$.  Suppose that we choose a
projection dimension of the form $m = \CONLS \, \alpha \usedim$, where
the control parameter $\alpha$ ranged over the interval $[0,1]$.
Corollary~\ref{CorUncLS} predicts that the approximation error should
converge to $1$ under this scaling, for each choice of $\numobs$.

\begin{figure}[h!]
\begin{center}
\widgraph{\MYTEXTWIDTH}{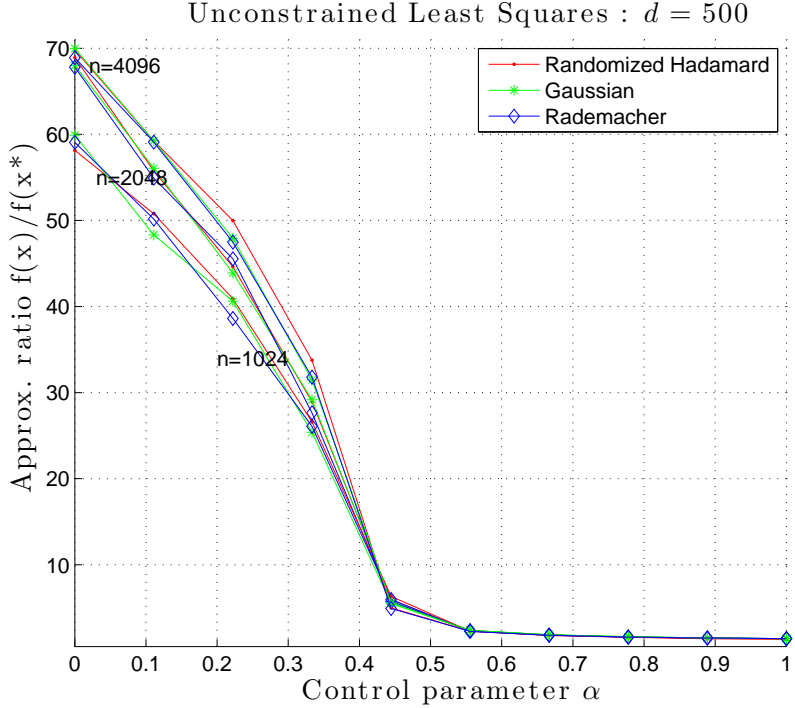}
  \caption{Comparison of Gaussian, Rademacher and randomized Hadamard
    sketches for unconstrained least squares.  Each curve plots the
    approximation ratio $f(\xhat)/f(\xstar)$ versus the control
    parameter $\alpha$, averaged over $T_{trial} = 100$ trials, for projection
    dimensions $\numproj = \CONLS \alpha \usedim$ and for problem
    dimensions $\usedim = 500$ and $\numobs \in \{1024,2048,4096\}$.}
\label{FigUncLS}
\end{center}
\end{figure}

Figure~\ref{FigUncLS} shows the results of these experiments, plotting
the approximation ratio $f(\xhat)/f(\xstar)$ versus the control
parameter $\alpha$. Consistent with Corollary~\ref{CorUncLS},
regardless of the choice of $\numobs$, once the projection dimension
is a suitably large multiple of $\rank(\Amat) = 500$, the
approximation quality becomes very good.


\subsection{$\ell_1$-constrained least squares}
\label{SecLasso}

We now turn a constrained form of least-squares, in which the geometry
of the tangent cone enters in a more interesting way.  In particular,
consider the following $\ell_1$-constrained least squares program,
known as the Lasso~\cite{Chen98,Tibshirani96}
\begin{align}
\label{EqnDefnLasso}
\xstar & \in \arg \min_{\|x\|_1 \leq R} \|\Amat x-y\|^2_2.
\end{align}
It is is widely used in signal processing and statistics for sparse
signal recovery and approximation.  

In this section, we show that as a corollary of Theorem~\ref{ThmMain},
this quadratic program can be sketched logarithmically in dimension
$\usedim$ when the optimal solution to the original problem is sparse.
In particular, assuming that $\xstar$ is unique, we let $\kdim$ denote
the number of non-zero coefficients of the unique solution to the
above program. (When $\xstar$ is not unique, we let $\kdim$ denote the
minimal cardinality among all optimal vectors). Define the $\ell_1$-restricted eigenvalues of the given
data matrix $\Amat$ as
\begin{align} 
\label{EqnDefnRE}
\sigkminsq(\Amat) \defn \min_{ \substack{\|z\|_2 = 1 \\\|z\|_1\le
    2\sqrt{k}}} \|Az\|^2_2, \quad \mbox{and} \quad \sigkmaxsq(\Amat)
\defn \max_{ \substack{\|z\|_2 = 1 \\\|z\|_1\le 2\sqrt{k}}}
\|Az\|^2_2.
\end{align}
\bcors[Approximation guarantees for $\ell_1$-constrained least
  squares]
\label{CorLasso}
Consider the $\ell_1$-constrained least squares problem~\eqref{EqnDefnLasso}:
\begin{enumerate}
\item[(a)] For sub-Gaussian sketches, a sketch dimension lower bounded
  by
\begin{align}
\label{EqnSubgaussLasso}
\numproj & \geq \frac{\UNICON_0}{\MYEPS^2} \; \min \biggr
\{\rank(\Amat), \; \max \limits_{j= 1, \ldots ,\usedim }
\frac{\|a_j\|_2^2}{\sigkminsq(\Amat)} \; \kdim \log(\usedim) \, \biggr
\}
\end{align}
guarantees that the sketched solution is
$\MYEPS$-optimal~\eqref{EqnDefnEpsOptimal} with probability at least
$1 - c_1 \CEXP{-c_2 \numproj \delta^2}$.
\item[(b)] For ROS sketches, a sketch dimension lower bounded by
\begin{align}
\label{EqnROSLasso}
\numproj & > \frac{\UNICON'_0}{\MYEPS^2} \log^4(\numobs) \min \biggr
\{ \rank(A), \; \frac{\big(\frac{\max_j\|a_j\|_2^2}{\sigkminsq(\Amat)} \kdim
  \log(\usedim) \big)^2}{\log^4(\numobs)}, \;
\big(\frac{\sigkmaxsq(\Amat)}{\sigkminsq(\Amat)} \big)^2 \, \kdim
\log(\usedim) \biggr \}
\end{align}
guarantees that the sketched solution is
$\MYEPS$-optimal~\eqref{EqnDefnEpsOptimal} with probability at least
$1 - c_1 \CEXP{-c_2 \numproj \delta^2}$.
\end{enumerate}
\ecors

\noindent We note that part (a) of this corollary improves the result
of Zhou et al.~\cite{Zhou09}, which establishes consistency of Lasso
with a Gaussian sketch dimension of the order $\kdim^2 \log(\usedim
\numobs \kdim)$, in contrast to the $\kdim \log(\usedim)$ requirement
in the bound~\eqref{EqnSubgaussLasso}.  To be more precise, these two
results are slightly different, in that the result~\cite{Zhou09}
focuses on support recovery, whereas Corollary~\ref{CorLasso}
guarantees a $\delta$-accurate approximation of the cost function.

Let us consider the complexity of solving the sketched problem using
different methods.  In the regime $\numobs > \usedim$, the complexity
of solving the original Lasso problem as a linearly constrained
quadratic program via interior point solvers is
$\order(\numobs\usedim^2)$ per iteration (e.g., see Nesterov and
Nemirovski~\cite{NesNem94}).  Thus, computing the sketched data and
solving the sketched Lasso problem requires
$\order(\numobs\usedim\numproj+\numproj\usedim^2)$ operations for
sub-Gaussian sketches, and
$\order(\numobs\usedim\log(\numproj)+\numproj\usedim^2)$ for ROS
sketches.

Another popular choice for solving the Lasso problem is to use a
first-order algorithm~\cite{Nesterov07}; such algorithms require
$\order(\numobs\usedim)$ operations per iteration, and yield a solution
that is $\order(1/T)$-optimal within $T$ iterations.  If we apply such
an algorithm to the sketched version for $T$ steps, then we obtain a
vector such that
\begin{align*}
f(\xhat) & \leq (1 + \MYEPS)^2 f(\xstar) + \order(\frac{1}{T}).
\end{align*}
Overall, obtaining this guarantee requires
$\order(\numobs\usedim\numproj+\numproj\usedim T)$ operations for
sub-Gaussian sketches, and
$\order(\numobs\usedim\log(\numproj)+\numproj\usedim T)$ operations
for ROS sketches.

\HACKPROOF Let $\Sset$ denote the support of the optimal solution
$x^*$.  The tangent cone to the $\ell_1$-norm constraint at the
optimum $x^*$ takes the form
\begin{align}
\label{EqnEllOneTangentCone}
\ConeSet & = \big \{\Delta \in \real^\usedim \, \mid \,
\inprod{\Delta_\Sset}{ \zhat_\Sset} + \|\Delta_{\Sbar}\|_1 \leq 0 \},
\end{align}
where $\zhat_\Sset \defn \sign(\xstar_\Sset) \in \{-1, +1\}^\kdim$ is
the sign vector of the optimal solution on its support.  By the
triangle inequality, any vector $\Delta \in \ConeSet$ satisfies the
inequality
\begin{align}
\|\Delta\|_1 \, \leq \, 2 \|\Delta_\Sset\|_1 \leq 2 \sqrt{\kdim}
\|\Delta_S\|_2 \; \leq \; 2 \sqrt{\kdim} \|\Delta\|_2. \label{EqnL1StarShaped}
\end{align}
If $\|A \Delta\|_2 = 1$, then by the definition~\eqref{EqnDefnRE}, we
also have the upper bound $\|\Delta\|_2 \leq
\frac{1}{\sqrt{\sigkminsq(\Amat)}}$, whence
\begin{align}
\label{EqnHanaSleep}
\inprod{\Amat \Delta}{g} & \leq 2 \sqrt{|\Sset|} \; \|\Delta\|_2
\|\Amat^T g\|_\infty \; \leq \frac{2 \sqrt{|\Sset|} \, \|\Amat^T
  g\|_\infty}{\sqrt{\sigkminsq(\Amat)}}.
\end{align}
Note that $\Amat^T g$ is a $\usedim$-dimensional Gaussian vector, in
which the $j^{th}$-entry has variance $\|a_j\|^2_2$.  Consequently,
inequality~\eqref{EqnHanaSleep} combined with standard Gaussian tail
bounds~\cite{LedTal91} imply that
\begin{align}
\SUPERWIDTH & \leq 6 \sqrt{\kdim \, \log(\usedim )} \max_{j =1,
  \ldots, \usedim} \frac{\|a_j\|_2}{\sqrt{\sigkminsq(\Amat)}}.
\end{align}
Combined with the bound from Corollary~\ref{CorUncLS}, also applicable
in this setting, the claim~\eqref{EqnSubgaussLasso} follows.

Turning to part (b), the first lower bound involving $\rank(A)$
follows from Corollary~\ref{CorUncLS}.  The second lower bound follows
as a corollary of Theorem~\ref{ThmHadamard} in application to the
Lasso; see Appendix~\ref{AppCorROSLasso} for the calculations.  The
third lower bound follows by a specialized argument given in
Section~\ref{SecEllOneSharpen}.

\HACKENDPROOF

In order to investigate the prediction of Corollary~\ref{CorLasso}, we
generated a random ensemble of sparse linear regression problems as
follows.  We first generated a data matrix $\Amat \in \real^{4096
  \times 500}$ by sampling i.i.d. standard Gaussian entries, and then
a $k'$-sparse base vector $x_0 \in \real^\usedim$ by choosing a
uniformly random subset $\Sset$ of size $k' = \usedim/10$, and setting its
entries to in $\{-1, +1\}$ independent and equiprobably.  Finally, we
formed the data vector $y = \Amat x_0 + w$, where the noise vector $w
\in \real^\numobs$ has i.i.d. $N(0, \nu^2)$ entries.

\begin{figure}[h!]
\begin{center}
  \widgraph{\MYTEXTWIDTH}{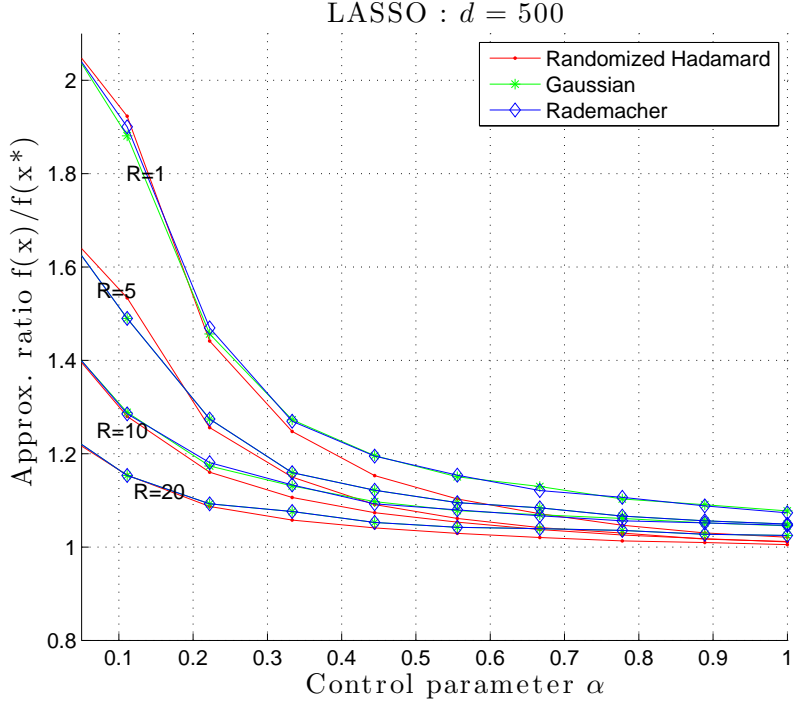}
  \caption{Comparison of Gaussian, Rademacher and randomized Hadamard
    sketches for the Lasso program~\eqref{EqnDefnLasso}.  Each curve
    plots the approximation ratio $f(\xhat)/f(\xstar)$ versus the
    control parameter $\alpha$, averaged over $T_{trial} = 100$ trials, for
    projection dimensions $\numproj = \CONLAS \alpha \|\xstar\|_0 \log
    \usedim$, problem dimensions $(\numobs, \usedim) = (4096, 500)$,
    and $\ell_1$-constraint radius $R \in \{1,5,10,20\}$.}
\label{FigLasso}
\end{center}
\end{figure}

In our experiments, we solved the Lasso~\eqref{EqnDefnLasso} with a
choice of radius parameter $R \in \{1,5,10,20\}$, and set $\kdim =
\|\xstar\|_0$.  We then set the projection dimension $\numproj =
\CONLAS \kdim \log \usedim$ where $\alpha \in (0,1)$ is a control
parameter, and solved the sketched Lasso for Gaussian, Rademacher and
randomized Hadamard sketching matrices.  Our theory predicts that the
approximation ratio should tend to one as the control parameter
$\alpha$ increases.  The results are plotted in Figure~\ref{FigLasso},
and confirm this qualitative prediction.


\subsection{Compressed sensing and noise folding} 

It is worth noting that various compressed sensing results can be
recovered as a special case of Corollary~\ref{CorLasso}---more
precisely, one in which the ``data matrix'' $\Amat$ is simply the
identity (so that $\numobs = \usedim$).  With this choice, the
original problem~\eqref{EqnOriginalProblem} corresponds to the
classical denoising problem, namely
\begin{align}
\label{EqnDenoise}
\xstar & = \arg \min_{x \in \Constraint} \|x-y\|^2_2,
\end{align}
so that the cost function is simply $f(x) = \|x - y\|_2^2$.  With the
choice of constraint set $\Constraint = \{\|x\|_1 \leq R \}$, the
optimal solution $\xstar$ to the original problem is unique, and can
be obtained by performing a coordinate-wise soft-thresholding
operation on the data vector $y$.  For this choice, the sketched
version of the de-noising problem~\eqref{EqnDenoise} is given by
\begin{align}
\label{EqnCSRecovery}
\xhat & = \arg \min_{x \in \Constraint} \| \Sketch x- \Sketch y\|_2^2
\end{align}

\paragraph{Noiseless version:}
In the noiseless version of compressed sensing, we have $y = \bar{x}
\in \Constraint$, and hence the optimal solution to the original
``denoising'' problem~\eqref{EqnDenoise} is given by $\xstar =
\bar{x}$, with optimal value
\begin{align*}
f(\xstar) = \|\xstar - \bar{x}\|_2^2= 0.
\end{align*}
Using the sketched data vector $\Sketch \bar{x} \in \real^\numproj$,
we can solve the sketched program~\eqref{EqnCSRecovery}.  If doing so
yields a $\MYEPS$-approximation $\xhat$, then in this special case, we
are guaranteed that
\begin{align}
\label{EqnExactRecovery}
\|\xhat - \bar{x}\|_2^2 = f(\xhat) \; \leq \; (1 + \delta)^2 f(\xstar)
\; = \; 0,
\end{align}
which implies that we have exact recovery---that is, $\xhat =
\bar{x}$.

\paragraph{Noisy versions:}   In a more general setting,
we observe the vector $y = \bar{x} + w$, where $\bar{x} \in
\Constraint$ and $w \in \real^\numobs$ is some type of observation
noise.  The sketched observation model then takes the form
\begin{align*}
\Sketch y & = \Sketch \bar{x} + \Sketch w,
\end{align*}
so that the sketching matrix is applied to both the true vector
$\bar{x}$ and the noise vector $w$.  This set-up corresponds to an
instance of compressed sensing with ``folded'' noise (e.g., see the
papers~\cite{CastroEldar11,Aeron10}), which some argue is a more
realistic set-up for compressed sensing.  In this context, our results
imply that the sketched version satisfies the bound
\begin{align}
\label{EqnApproximateRecovery}
\|\xhat - y\|_2^2 & \leq \big(1 + \MYEPS \big)^2 \, \| \xstar -
y\|_2^2.
\end{align}
If we think of $y$ as an approximately sparse vector and $\xstar$ as
the best approximation to $y$ from the $\ell_1$-ball, then this
bound~\eqref{EqnApproximateRecovery} guarantees that we recover a
$\MYEPS$-approximation to the best sparse approximation.  Moreover,
this bound shows that the compressed sensing error should be closely
related to the error in denoising, as has been made precise in recent
work~\cite{Donoho13}. \\

\noindent Let us summarize these conclusions in a corollary: \\
\bcors
\label{CorCompressed}
Consider an instance of the denoising problem~\eqref{EqnDenoise} when
$\Constraint = \{x \in \real^\numobs \, \mid \, \|x\|_1 \leq R \}$.
\begin{enumerate}
\item[(a)] For sub-Gaussian sketches with projection dimension
  {$\numproj \geq \frac{\UNICON_0}{\MYEPS^2} \, \|\xstar\|_0 \log
    \usedim$}, we are guaranteed exact recovery in the noiseless
  case~\eqref{EqnExactRecovery}, and $\MYEPS$-approximate
  recovery~\eqref{EqnApproximateRecovery} in the noisy case, both with
  probability at least $\HACKPROB$.

\item[(b)] For ROS sketches, the same conclusions hold with
  probability $\ROSHACKN$ using a sketch dimension 
\begin{align}
\numproj \geq \frac{\UNICON_0}{\MYEPS^2} \, \min \big \{ \|\xstar\|_0
\log^5 \usedim , \|\xstar\|^2_0 \log \usedim \big \}.
\end{align}
\end{enumerate}
\ecors

Of course, a more general version of this corollary holds for any
convex constraint set $\Constraint$, involving the Gaussian/Rademacher
width functions.  In this more setting, the corollary generalizes
results by Chandrasekaran et al.~\cite{ChaRec12}, who studied
randomized Gaussian sketches in application to atomic norms, to other
types of sketching matrices and other types of constraints.  They
provide a number of calculations of widths for various atomic norm
constraint sets, including permutation and orthogonal matrices, and
cut polytopes, which can be used in conjunction with the more general
form of Corollary~\ref{CorCompressed}.


\subsection{Support vector machine classification}

Our theory also has applications to learning linear classifiers based
on labeled samples.  In the context of binary classification, a
labeled sample is a pair $(a_i, z_i)$, where the vector $a_i \in
\real^\numobs$ represents a collection of features, and $z_i \in \{-1,
+1\}$ is the associated class label.  A linear classifier is specified
by a function $a \mapsto \sign(\inprod{w}{a}) \in \{-1, +1\}$, where
$w \in \real^\numobs$ is a weight vector to be estimated.

Given a set of labelled patterns $\{a_i,z_i\}_{i=1}^\usedim$, the
support vector machine~\cite{Cristianini00,SteChr08} estimates the weight vector $w^*$ by minimizing the function
\begin{align}
w^* & = \arg \min_{w \in \real^\numobs} \Big \{ \frac{1}{2C} \sum_{i=1}^\usedim
g(y_i, \inprod{w}{a_i}) + \frac{1}{2} \|w\|_2^2 \Big \}.
\end{align}
In this formulation, the squared hinge loss $g(w) \defn
(1-y_i\inprod{w}{a_i})_{+}^2$ is used to measure the performance of
the classifier on sample $i$, and the quadratic penalty $\|w\|_2^2$
serves as a form of regularization.

By considering the dual of this problem, we arrive at a least-squares
problem that is amenable to our sketching techniques.  Let $A \in
\real^{\numobs \times \usedim}$ be a matrix with $a_i \in
\real^\numobs$ as its $i^{th}$ column, and let $D = \diag(z) \in
\real^{\usedim \times \usedim}$ be a diagonal matrix and let $B^T = [(\Amat D)^T ~\frac{1}{C}I]$.
 With this
notation, the associated dual problem (e.g. see the paper~\cite{Li09})
takes the form
\begin{align}
\label{EqnSVMDual}
\xstar & \defn \arg \min_{x \in \real^\usedim} \|B x\|^2_2
\qquad \mbox{such that $x \geq 0$ and $\sum \limits_{i=1}^\usedim x_i
  = 1$.}
\end{align}
The optimal solution $\xstar \in \real^\usedim$ corresponds to a
vector of weights associated with the samples: it specifies the
optimal SVM weight vector via $w^* = \sum_{i=1}^\usedim x^*_i z_i
a_i$.  It is often the case that the dual solution $\xstar$ has
relatively few non-zero coefficients, corresponding to samples that
lie on the so-called margin of the support vector machine.

The sketched version is then given by
\begin{align}
\label{EqnSVMDualSketched}
\xhat & \defn \arg \min_{x \in \real^\usedim} \|\Sketch B
x\|^2_2 \qquad \mbox{such that $x \geq 0$ and $\sum
  \limits_{i=1}^\usedim x_i = 1$.}
\end{align}
The simplex constraint in the quadratic program~\eqref{EqnSVMDual},
although not identical to an $\ell_1$-constraint, leads to similar
scaling in terms of the sketch dimension.

\bcors[Sketch dimensions for support vector machines]
\label{CorSVM}
Given a collection of labeled samples $\{(a_i, z_i)\}_{i=1}^\usedim$,
let $\|\xstar\|_0$ denote the number of samples on the margin in the
SVM solution~\eqref{EqnSVMDual}.  Then given a sub-Gaussian sketch
with dimension
\begin{align}
\label{EqnSVMguarantee}
\numproj & \geq \frac{c_0}{\MYEPS^2} \, \|\xstar\|_0 \; \log(\usedim)
\max_{j=1, \ldots, \usedim} \frac{\|a_j\|^2_2}{\sigkminsq(\Amat)},
\end{align}
the sketched solution~\eqref{EqnSVMDualSketched} is $\MYEPS$-optimal
with probability at least $\HACKPROB$.  
\ecors
\noindent We omit the proof, as the calculations specializing from
Theorem~\ref{ThmMain} are essentially the same as those of
Corollary~\ref{CorLasso}. The computational complexity of solving the SVM problem as a linearly constrained quadratic problem is same with the Lasso problem, hence same conclusions apply. \\

\begin{figure}[h!]
\begin{center}
\widgraph{\MYTEXTWIDTH}{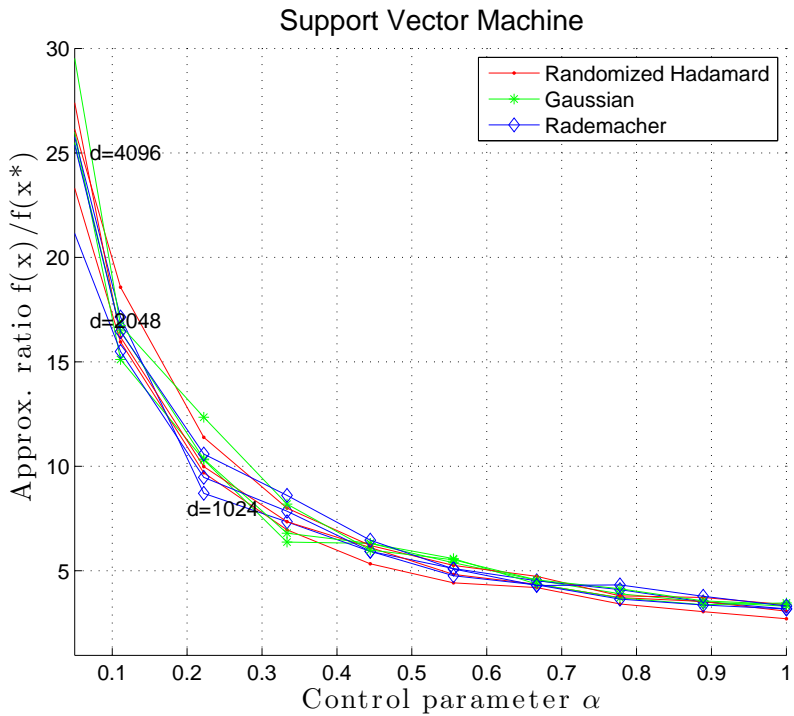}
\caption{Comparison of Gaussian, Rademacher and randomized Hadamard
  sketches for the support vector machine~\eqref{EqnSVMDual}.  Each
  curve plots the approximation ratio $f(\xhat)/f(\xstar)$ versus the
  control parameter $\alpha$, averaged over $T_{trial} = 100$ trials, for
  projection dimensions $\numproj = \CONSVM \alpha \| \xstar\|_0 \log
  \usedim$, and problem dimensions $\usedim \in \{1024, 2048,
  4096\}$.}
\label{FigSVM}
\end{center}
\end{figure}

In order to study the prediction of Corollary~\ref{CorSVM}, we
generated some classification experiments, and tested the performance
of the sketching procedure.  Consider a two-component Gaussian mixture
model, based on the component distributions $N(\mu_0, I)$ and
$N(\mu_1, I)$, where $\mu_0$ and $\mu_1$ are uniformly distributed in
$[-3,3]$.  Placing equal weights on each component, we draw $\usedim$
samples from this mixture distribution, and then use the resulting
data to solve the SVN dual program~\eqref{EqnSVMDual}, thereby
obtaining an optimal linear decision boundary specified by the vector
$x^*$.  The number of non-zero entries $\|\xstar\|_0$ corresponds to
the number of examples on the decision boundary, known as support
vectors.  We then solve the sketched
version~\eqref{EqnSVMDualSketched}, using either Gaussian, Rademacher
or randomized Hadamard sketches, and using a projection dimension
scaling as $\numproj = \CONSVM \, \alpha \|\xstar\|_0 \log \usedim$,
where $\alpha \in [0,1]$ is a control parameter.  We repeat this
experiment for problem dimensions $\usedim \in \{1024, 2048, 4096 \}$,
performing $T_{trial} = 100$ trials for each choice of $(\alpha,
\usedim)$.

Figure~\ref{FigSVM} shows plots of the approximation ratio versus the
control parameter.  Each bundle of curves corresponds to a different
problem dimension, and has three curves for the three different sketch
types.  Consistent with the theory, in all cases, the approximation
error approaches one as $\alpha$ scales upwards.


It is worthwhile noting that similar sketching techniques can be
applied to other optimization problems that involve the unit simplex
as a constraint.  Another instance is the Markowitz formulation of the
portfolio optimization problem~\cite{Markowitz59}.  Here the goal is
to estimate a vector $x \in \real^\usedim$ in the unit simplex,
corresponding to non-negative weights associated with each of
$\usedim$ possible assets, so as to minimize the variance of the
return subject to a lower bound on the expected return.  More
precisely, we let $\mu \in \real^\usedim$ denote a vector
corresponding to mean return associated with the assets, and we let
$\Sigma \in \real^{\usedim \times \usedim}$ be a symmetric, positive
semidefinite matrix, corresponding to the covariance of the returns.
Typically, the mean vector and covariance matrix are estimated from
data.  Given the pair $(\mu, \Sigma)$, the Markowitz allocation is
given by
\begin{align} 
\xstar & = \arg \min_{x \in \real^\usedim} x^T \Sigma x \qquad
\mbox{such that $\inprod{\mu}{x} \geq \gamma$, $x \geq 0$ and
  $\sum_{j=1}^\usedim x_j = 1$.}
\end{align}
Note that this problem can be written in the same form as the SVM,
since the covariance matrix $\Sigma \succeq 0$ can be factorized as
$\Sigma = \Amat^T \Amat$.  Whenever the expected return constraint
$\inprod{\mu}{x} \geq \gamma$ is active at the solution, the tangent
cone is given by
\begin{align*}
\ConeSet & = \big \{ \Delta \in \real^\usedim \, \mid \,
\inprod{\mu}{\Delta} \geq 0, \quad \sum_{j=1}^\usedim \Delta_j =0,
\quad \Delta_{S^c} \geq 0 \}
\end{align*}
where $S$ is the support of $\xstar$.  This tangent cone is a subset
of the tangent cone for the SVM, and hence the bounds of
Corollary~\ref{CorSVM} also apply to the portfolio optimization
problem.


\subsection{Matrix estimation with nuclear norm regularization}

We now turn to the use of sketching for matrix estimation problems,
and in particular those that involve nuclear norm constraints.  Let
$\Constraint \subset \real^{\usedim_1 \times \usedim_2}$ be a convex
subset of the space of all $\usedim_1 \times \usedim_2$ matrices.
Many matrix estimation problems can be written in the general form
\begin{align*}
\min_{X \in \Constraint} \|y - \Aop(X)\|_2^2
\end{align*}
where $y \in \real^{\numobs}$ is a data vector, and $\Aop$ is a linear
operator from $\real^{\usedim_1 \times \usedim_2}$ to
$\real^{\numobs}$.  Letting $\myvec$ denote the vectorized form of a
matrix, we can write $\Aop(X) = \Amat \myvec(X)$ for a suitably
defined matrix $\Amat \in \real^{\numobs \times D}$, where $D =
\usedim_1 \usedim_2$.  Consequently, our general sketching techniques
are again applicable.

In many matrix estimation problems, of primary interest are matrices
of relatively low rank.  Since rank constraints are typically
computationally intractable, a standard convex surrogate is the
nuclear norm of matrix, given by the sum of its singular values
\begin{align}
\nucnorm{X} & = \sum_{j=1}^{\min \{\usedim_1, \usedim_2\}}
\sigma_j(X).
\end{align}

As an illustrative example, let us consider the problem of weighted
low-rank matrix approximation, Suppose that we wish to approximate a
given matrix $Z \in \real^{\usedim_1 \times \usedim_2}$ by a low-rank
matrix $X$ of the same dimensions, where we measure the quality of
approximation using a weighted Frobenius norm
\begin{align}
\label{EqnWeightedFrobenius}
\matsnorm{Z - X}{\omega}^2 & = \sum_{j = 1}^{\usedim_2} \omega^2_j
\|z_j - x_j\|_2^2,
\end{align}
where $z_j$ and $x_j$ are the $j^{th}$ columns of $Z$ and $X$
respectively, and $\omega \in \real^{\usedim_2}$ is a vector of
non-negative weights.  If the weight vector is uniform ($\omega_j = c$
for all $j = 1, \ldots, \usedim$), then the norm
$\matsnorm{\cdot}{\omega}$ is simply the usual Frobenius norm, a
low-rank minimizer can be obtained by computing a partial singular
value decomposition of the data matrix $Y$.  For non-uniform weights,
it is no longer easy to solve the rank-constrained minimization
problem.  Accordingly, it is natural to consider the convex relaxation
\begin{align}
\label{EqnOriginalWeightedLowRank}
\Xstar & \defn \arg \min_{\nucnorm{X} \leq R} \matsnorm{Z -
  X}{\omega}^2,
\end{align}
in which the rank constraint is replaced by the nuclear norm
constraint $\nucnorm{X} \leq R$.  This program can be written in an
equivalent vectorized form in dimension $D = \usedim_1 \usedim_2$ by
defining the block-diagonal matrix $\Amat =
\operatorname{blkdiag}(\omega_1 I, \ldots, \omega_{\usedim_2} I)$, as
well as the vector $y \in \real^{D}$ whose $j^{th}$ block is given by
$\omega_j y_j$.  We can then consider the equivalent problem $\Xstar
\defn \arg \min \limits_{\nucnorm{X} \leq R} \|y - \Amat
\myvec(X)\|_2^2$, as well as its sketched version
\begin{align}
\label{EqnWeightedLowRankSketch}
\Xhat & \defn \arg \min_{\nucnorm{X} \leq R} \|\Sketch y - \Sketch
\Amat \myvec(X)\|_2^2.
\end{align}

Suppose that the original optimum $\Xstar$ has rank $\rdim$: it then
be described using at $\order(\rdim (\usedim_1 + \usedim_2))$ real
numbers. Intuitively, it should be possible to project the original
problem down to this dimension while still guaranteeing an accurate
solution.  The following corollary provides a rigorous confirmation of
this intuition: \\
\bcors[Sketch dimensions for weighted low-rank approximation]
\label{CorNuclear}
Consider the weighted low-rank approximation
problem~\eqref{EqnOriginalWeightedLowRank} based on a weight vector
with condition number $\kappa^2(\omega) = \frac{\max \limits_{j = 1,
    \ldots, \usedim} \omega_j^2}{\min \limits_{j = 1, \ldots, \usedim}
  \omega_j^2}$, and suppose that the optimal solution has rank $\rdim
= \rank(\Xstar)$.
\begin{enumerate}
\item[(a)] For sub-Gaussian sketches, a sketch dimension lower bounded
  by
\begin{align}
\label{EqnSubgaussNuclear}
\numproj & \geq \frac{\UNICON_0}{\MYEPS^2} \; \kappa^2(\omega) \;
\rdim \, (\usedim_1 + \usedim_2)
\end{align}
guarantees that the sketched solution~\eqref{EqnWeightedLowRankSketch}
is $\MYEPS$-optimal~\eqref{EqnDefnEpsOptimal} with probability at
least \mbox{$\HACKPROB$.}
\item[(b)] For ROS sketches, a sketch dimension lower bounded by
\begin{align}
\label{EqnROSNuclear}
\numproj & > \frac{\UNICON'_0}{\MYEPS^2} \kappa^2(\omega) \rdim \,
(\usedim_1 + \usedim_2) \; \log^4(\usedim_1 \usedim_2).
\end{align}
guarantees that the sketched solution~\eqref{EqnWeightedLowRankSketch}
is $\MYEPS$-optimal~\eqref{EqnDefnEpsOptimal} with probability at
least \mbox{$\HACKPROB$}.
\end{enumerate}
\ecors
\noindent For this particular application, the use of sketching is not
likely to lead to substantial computational savings, since the
optimization space remains $\usedima \usedimb$ dimensional in both the
original and sketched versions. However, the lower dimensional nature
of the sketched data can be still very useful in reducing storage
requirements and privacy-sensitive optimization.
\HACKPROOF

We prove part (a) here, leaving the proof of part (b) to
Section~\ref{SecNuclearSharpen}.  Throughout the proof, we adopt the
shorthand notation $\ommin = \min \limits_{j = 1, \ldots, \usedim}
\omega_j$ and $\ommax = \max \limits_{j = 1, \ldots, \usedim}
\omega_j$.  As shown in past work on nuclear norm regularization (see
Lemma 1 in the paper~\cite{NegWai09}), the tangent cone of the nuclear
norm constraint $\nucnorm{X} \leq R$ at a rank $\rdim$ matrix is
contained within the cone
\begin{align}
\ConeSet' & = \big \{ \Delta \in \real^{\usedim_1 \times \usedim_2} \,
\mid \, \nucnorm{\Delta} \leq 2 \sqrt{\rdim} \frobnorm{\Delta} \big
\}.
\end{align}
For any matrix $\Delta$ with $\|A \myvec(\Delta)\|_2 = 1$, we must
have $\frobnorm{\Delta} = \|\myvec(\Delta)\|_2 \leq \frac{1}{\ommin}$.
By definition of the Gaussian width, we then have
\begin{align*}
\SUPERWIDTH & \leq \frac{1}{\ommin} \Exs \big[ \sup_{
    \substack{\nucnorm{\Delta} \leq 2 \sqrt{\rdim}}} |\inprod{A^T g}{
    \myvec(\Delta)}|\big].
\end{align*}
Since $A^T$ is a diagonal matrix, the vector $A^T g$ has independent
entries with maximal variance $\ommax^2$.  Letting $G \in
\real^{\usedim_1 \times \usedim_2}$ denote the matrix formed by
segmenting the vector $A^T g$ into $\usedim_2$ blocks of length
$\usedim_1$, we have
\begin{align*}
\SUPERWIDTH & \leq \frac{1}{\ommin} \Exs \big[ \sup_{
    \substack{\nucnorm{\Delta} \leq 2 \sqrt{\rdim}}} |\trace(G
  \Delta)| \big] \; \leq \; \frac{2 \sqrt{\rdim}}{\ommin} \, \Exs
\big[ \opnorm{G} \big]
\end{align*}
where we have used the duality between the operator and nuclear norms.
By standard results on operator norms of Gaussian random
matrices~\cite{DavSza01}, we have
$\Exs[\opnorm{G}]  \leq \ommax \big( \sqrt{\usedima} + \sqrt{\usedimb}
\big)$, and hence
\begin{align*}
\SUPERWIDTH & \leq 2 \frac{\ommax}{\ommin} \, \sqrt{\rdim} \, \big(
\sqrt{\usedima} + \sqrt{\usedimb} \big).
\end{align*}
Thus, the bound~\eqref{EqnSubgaussNuclear} follows as a corollary of
Theorem~\ref{ThmMain}.
 \HACKENDPROOF

%

\subsection{Group sparse regularization}

As a final example, let us consider optimization problems that involve
constraints to enforce group sparsity.  This notion is a
generalization of elementwise sparsity, defined in terms of a
partition $\GroupSet$ of the index set $[\usedim] = \{1, 2, \ldots,
\usedim \}$ into a collection of non-overlapping subsets, referred to
as groups.  Given a group $g \in \GroupSet$ and a vector $x \in
\real^\usedim$, we use $x_g \in \real^{|g|}$ to denote the sub-vector
indexed by elements of $g$.  A basic form of the group Lasso
norm~\cite{YuaLi06} is given by
\begin{align}
\label{EqnGroupLassoNorm}
\|x\|_\GroupSet & = \sum_{g \in \GroupSet} \|x_g\|_2.
\end{align}
Note that in the special case that $\GroupSet$ consists of $\usedim$
groups, each of size $1$, this norm reduces to the usual
$\ell_1$-norm.  More generally, with non-trivial grouping, it defines
a second-order cone constraint~\cite{Boyd02}.  Bach et
al.~\cite{BacJenMaiObo12b} provide an overview of the group Lasso
norm~\eqref{EqnGroupLassoNorm}, as well as more exotic choices for
enforcing group sparsity.

Here let us consider the problem of sketching the second-order cone
program (SOCP)
\begin{align}
\label{EqnGroupLassoProgram}
\xstar & = \arg \min_{\|x \|_\GroupSet \leq R} \| \Amat x- \yvec
\|_2^2.
\end{align}
We let $\kdim$ denote the number of active groups in the optimal
solution $\xstar$---that is, the number of groups for which
$\xstar_g\neq 0$.  For any group $\group \in \GroupSet$, we use
$\Amat_\group$ to denote the $\numobs \times |\group|$ sub-matrix with
columns indexed by $\group$.  In analogy to the sparse RE
condition~\eqref{EqnDefnRE}, we define the group-sparse restricted
eigenvalue \mbox{$\siggroupminsq(\Amat) \defn \min_{ \substack{\|z\|_2
      = 1 \\ \|z\|_\GroupSet \leq 2\sqrt{\kdim}}} \|Az\|^2_2$.}

\bcors[Guarantees for group-sparse least-squares squares]
\label{CorGroupLasso}
For the group Lasso program~\eqref{EqnGroupLassoProgram} with maximum
group size $\gmax = \max_{g\in \GroupSet} |g|$, a
projection dimension lower bounded as
\begin{align}
\label{EqnSubgaussGroupLasso}
\numproj & \geq \frac{\UNICON_0}{\MYEPS^2} \; \min \biggr
\{\rank(\Amat), \; \max \limits_{\group \in \GroupSet}
\frac{\opnorm{A_\group}}{\siggroupminsq(\Amat)} \; \big (\kdim \log
|\GroupSet| + \kdim \gmax \big ) \, \biggr \}
\end{align}
guarantees that the sketched solution is
$\MYEPS$-optimal~\eqref{EqnDefnEpsOptimal} with probability at least
$\HACKPROB$.
\ecors 

\noindent Note that this is a generalization of
Corollary~\ref{CorLasso} on sketching the ordinary Lasso.  Indeed,
when we have $|\GroupSet| = \usedim$ groups, each of size $\gmax = 1$,
then the lower bound~\eqref{EqnSubgaussGroupLasso} reduces to the
lower bound~\eqref{EqnSubgaussLasso}.  As might be expected, the proof
of Corollary~\ref{CorGroupLasso} is similar to that of
Corollary~\ref{CorLasso}.  It makes use of some standard results on
the expected maxima of $\chi^2$-variates to upper bound the Gaussian
complexity; see the paper~\cite{NegRavWaiYu12} for more details on
this calculation.


\section{Proofs of main results}
\label{SecProofs}

We now turn to the proofs of our main results, namely
Theorem~\ref{ThmMain} on sub-Gaussian sketching, and
Theorem~\ref{ThmHadamard} on sketching with randomized orthogonal
systems.  At a high level, the proofs consists of two parts. The first
part is a deterministic argument, using convex optimality conditions.
The second step is probabilistic, and depends on the particular choice
of random sketching matrices.

\subsection{Main argument}

Central to the proofs of both Theorem~\ref{ThmMain}
and~\ref{ThmHadamard} are the following two variational quantities:
\begin{subequations}
\begin{align}
\label{EqnUsualZvar}
\ZINF & \defn \inf_{v \in \Amat \ConeSet \cap \SPHERE{\numobs}}
\frac{1}{\numproj} \| \Sketch v\|^2_2, \quad \mbox{and} \\
\label{EqnAnnoyingZvar}
\ZSUP & \defn \sup_{v \in \Amat \ConeSet \cap \SPHERE{\numobs}} \Big|
\inprod{\fixvec}{(\frac{\Sketch^T \Sketch}{\numproj} - I) \, v} \Big|,
\end{align}
\end{subequations}
where we recall that $\SPHERE{\numobs}$ is the Euclidean unit sphere
in $\real^\numobs$, and in equation~\eqref{EqnAnnoyingZvar}, the
vector $\fixvec \in \SPHERE{\numobs}$ is fixed but arbitrary.  These
are deterministic quantities for any fixed choice of sketching matrix
$\Sketch$, but random variables for randomized sketches.  The
following lemma demonstrates the significance of these two quantities:
\blems
\label{LemDeterministic}
For any sketching matrix $\Sketch \in \real^{\numproj \times
  \numobs}$, we have
\begin{align}
\label{EqnDeterministicClaim}
f(\xhat) & \leq \Big \{ 1 + 2 \,\frac{\ZSUPSTAR}{\ZINF} \Big\}^2 \;
f(\xstar)
\end{align}
\elems
\noindent Consequently, we see that in order to establish that $\xhat$
is $\MYEPS$-optimal, we need to control the ratio $\ZSUP/\ZINF$. \\
\HACKPROOF Define the error vector $\ehat \defn \xhat - \xstar$.  By
the triangle inequality, we have
\begin{align}
\label{EqnMertTriangle}
\|\Amat \xhat - \yvec\|_2 & \leq \|\Amat \xstar - \yvec\|_2 + \|\Amat
\ehat\|_2 \; = \; \|\Amat \xstar - \yvec\|_2 \, \big \{ 1 +
\frac{\|\Amat \ehat\|_2}{\|\Amat \xstar - \yvec\|_2} \big \}.
\end{align}
Squaring both sides yields
\begin{align*}
f(\xhat) & \leq \Big(1 + \frac{\|\Amat \ehat\|_2}{\|\Amat \xstar -
  \yvec\|_2} \Big)^2 \; f(\xstar).
\end{align*}
Consequently, it suffices to control the ratio $\frac{\|A
  \ehat\|_2}{\|A \xstar - \yvec\|_2}$, and we use convex optimality
conditions to do so.

Since $\xhat$ and $\xstar$ are optimal and feasible, respectively, for
the sketched problem~\eqref{EqnSketchedProblem}, we have $g(\xhat)
\leq g(\xstar)$, and hence (following some algebra)
\begin{align*}
\frac{1}{2} \|\Sketch \Amat \ehat\|_2^2 & \leq - \inprod{\Amat \xstar
  - \yvec }{(\Sketch^T \Sketch) \, \Amat \ehat} \\
& = - \inprod{\Amat \xstar - \yvec }{(\Sketch^T \Sketch-I) \, \Amat
  \ehat} - \inprod{\Amat \xstar - \yvec}{ \Amat \ehat},
\end{align*}
where we have added and subtracted terms.  Now by the optimality of
$\xstar$ for the original problem~\eqref{EqnOriginalProblem}, we have
\begin{align*}
\inprod{(\Amat \xstar - \yvec)}{\Amat \ehat} \; = \; \inprod{\Amat^T
  (\Amat \xstar - \yvec)}{\xhat - \xstar} \geq 0,
\end{align*}
and hence
\begin{align}
\frac{1}{2} \|S \Amat \ehat\|_2^2 & \leq \Big| \inprod{\Amat \xstar -
  \yvec }{(\Sketch^T \Sketch - I) \, \Amat \ehat} \Big|.
\end{align}
Renormalizing the right-hand side appropriately, we find that
\begin{align}
\label{EqnGoodBasic}
\frac{1}{2} \|\Sketch \Amat \ehat\|_2^2 & \leq \|\Amat \xstar - y \|_2
\, \|\Amat \ehat\|_2 \; \Big| \inprod{ \frac{\Amat \xstar -
    \yvec}{\|\Amat \xstar - \yvec \|_2 }}{(\Sketch^T \Sketch - I) \,
  \frac{A \ehat}{\|A \ehat\|_2}} \Big|.
\end{align}
By the optimality of $\xhat$, we have $\Amat \ehat \in \Amat
\ConeSet$, whence the basic inequality~\eqref{EqnGoodBasic} and
definitions~\eqref{EqnUsualZvar} and~\eqref{EqnAnnoyingZvar} imply
that
\begin{align*}
\frac{1}{2} \ZINF \, \|\Amat \ehat\|_2^2 & \leq \|\Amat \ehat\|_2 \;
\|\Amat \xstar - \yvec\|_2 \; \ZSUP
\end{align*}
Cancelling terms yields the inequality
\begin{align*}
\frac{\|\Amat \ehat\|_2}{\|\Amat \xstar - y\|_2} \leq 2 \,
\frac{\ZSUP}{\ZINF}.
\end{align*}
Combined with our earlier inequality~\eqref{EqnMertTriangle}, the
claim~\eqref{EqnDeterministicClaim} follows.
\HACKENDPROOF


\subsection{Proof of Theorem~\ref{ThmMain}}

In order to complete the proof of Theorem~\ref{ThmMain}, we need to
upper bound the ratio $\ZSUP/\ZINF$. The following lemmas provide such
control in the sub-Gaussian case.  As usual, we let $\Sketch \in
\real^{\numproj \times \numobs}$ denote the matrix with the vectors
$\{\sketch_i\}_{i=1}^\numproj$ as its rows.

\blems[Lower bound on $\ZINF$]
\label{LemZINF}
For i.i.d. $\sigma$-sub-Gaussian vectors
$\{\sketch_i\}_{i=1}^\numproj$, we have
\begin{align}
\underbrace{\inf_{v \in \Amat \ConeSet \cap \SPHERE{\numobs}}
  \frac{1}{\numproj} \| \Sketch v\|^2_2}_{\ZINF} & \geq 1 - \MYLEMEPS
\end{align}
with probability at least $1 - \exp \big(-\UNICON_1 \frac{\numproj
  \MYLEMEPS^2}{\sigma^4} \big)$.
\elems

\blems[Upper bound on $\ZSUP$]
\label{LemZSUP}
For i.i.d. $\sigma$-sub-Gaussian vectors
$\{\sketch_i\}_{i=1}^\numproj$ and any fixed vector $u \in
\SPHERE{\numobs}$, we have
\begin{align}
\underbrace{\sup_{v \in \Amat \ConeSet \cap \SPHERE{\numobs}} \Big|
  \inprod{\fixvec}{(\Sketch^T \Sketch - I) \, v} \Big|}_{\ZSUP} & \leq
\MYLEMEPS
\end{align}
with probability at least $1 - 6 \, \exp \big(-\UNICON_1
\frac{\numproj \MYLEMEPS^2}{\sigma^4} \big)$.
\elems

Taking these two lemmas as given, we can complete the proof of
Theorem~\ref{ThmMain}.  As long as $\delta \in (0,1/2)$, they imply
that
\begin{align}
2 \, \frac{\ZSUP}{\ZINF} & \leq \frac{2 \delta}{1- \delta} \; \leq \,
4 \delta \label{EqnZ1Z2Ratio}
\end{align}
with probability at least $1 - 4 \, \exp \big(-\UNICON_1
\frac{\numproj \MYLEMEPS^2}{\sigma^4} \big)$.  The rescaling $4 \delta
\mapsto \delta$, with appropriate changes of the universal constants,
yields the result.\\

It remains to prove the two lemmas.  In the sub-Gaussian case, both of
these results exploit a result due to Mendelson et
al.~\cite{MendelPajorTom07}: \\
\bprops
\label{PropMPT}
Let $\{\sketch_i\}_{i=1}^\numobs$ be i.i.d. samples from a zero-mean
$\sigma$-sub-Gaussian distribution with $\cov(\sketch_i) = I_{\numobs
  \times \numobs}$.  Then there are universal constants such that for
any subset $\YSET \subseteq \SPHERE{\numobs}$, we have
\begin{align}
\label{EqnMPT}
\sup_{y \in \YSET} \Big| y^T \big( \frac{\Sketch^T \Sketch}{\numproj}
- I_{\numobs \times \numobs} \big) y \Big| & \leq \UNICON_1
\frac{\Width(\YSET)}{\sqrt{\numproj}} + \delta
\end{align}
with probability at least $1 - \CEXP{- \frac{\UNICON_2 \numproj
    \delta^2}{\sigma^4}}$.
\eprops
\noindent This claim follows from their Theorem D, using the linear
functions $f_y(\sketch) = \inprod{\sketch}{y}$.  

\subsubsection{Proof of Lemma~\ref{LemZINF}}

Lemma~\ref{LemZINF} follows immediately from
Proposition~\ref{PropMPT}: in particular, the bound~\eqref{EqnMPT}
with the set $\YSET = \infset$ ensures that
\begin{align*}
\inf_{v \in \infset} \frac{\|\Sketch v\|_2^2}{\numproj} & \geq 1 -
\UNICON_1 \frac{\Width(\YSET)}{\sqrt{\numproj}} - \frac{\MYEPS}{2} \;
\stackrel{(i)}{\geq} \; 1 - \MYEPS,
\end{align*}
where inequality (i) follows as long as $\numproj >
\frac{\UNICON_0}{\MYEPS^2} \Width(\Amat \ConeSet)$ for a sufficiently
large universal constant.


\subsubsection{Proof of Lemma~\ref{LemZSUP}}

The proof of this claim is more involved.  Let us partition the set
$\VSET = \infset$ into two disjoint subsets, namely
\begin{align*}
\VSETPLUS = \{ v \in \VSET \, \mid \, \inprod{u}{v} \geq 0 \}, \quad
\mbox{and} \quad \VSET_{-} = \{ v \in \VSET \, \mid \, \inprod{u}{v} <
0 \}.
\end{align*}
Introducing the shorthand $Q = \frac{\Sketch^T \Sketch}{\numproj} -
I$, we then have 
\begin{align*}
\ZSUP \leq \sup \limits_{v \in \VSETPLUS} |u^T Q v| + \sup \limits_{v
  \in \VSETMINUS} |u^T Q v|,
\end{align*}
and we bound each of these terms in turn.\\

Beginning with the first term, for any $v \in \VSETPLUS$, the triangle
inequality implies that
\begin{align}
\label{EqnCorrectDecomposition}
|u^T Q v| & \leq \frac{1}{2} \big| (u + v)^T Q (u + v) \big| +
\frac{1}{2} \big| u^T Q u \big| + \frac{1}{2} \big| v^T Q v \big|.
\end{align}
Defining the set $\USETPLUS \defn \{ \frac{u + v}{\|u + v\|_2} \, \mid
\, v \in \VSETPLUS \}$, we apply Proposition~\ref{PropMPT} three times
in succession, with the choices $\YSET = \USETPLUS$, $\YSET =
\VSETPLUS$ and $\YSET = \{u\}$ respectively, which yields
\begin{subequations}
\begin{align}
\label{EqnHanaSmileOne}
\sup_{v \in \VSETPLUS} \frac{1}{\|u + v\|_2^2} \, \big| (u + v)^T Q (u
+ v) \big| & \leq \UNICON_1 \frac{\Width(\VSET)}{\sqrt{\numproj}} +
\delta \\
\label{EqnHanaSmileTwo}
\sup_{v \in \infset} \big| v^T Q v \big| & \leq \UNICON_1
\frac{\Width(\infset)}{\sqrt{\numproj}} + \delta, \qquad \mbox{and} \\
\label{EqnHanaSmileThree}
\big | u^T Q u \big| & \leq \UNICON_1
\frac{\Width(\{u\})}{\sqrt{\numproj}} + \delta.
\end{align}
\end{subequations}
All three bounds hold with probability at least $1 - 3 \CEXP{-
  \UNICON_2 \numproj \delta^2/\sigma^4}$.  Note that $\|u + v\|_2^2
\leq 4$, so that the bound~\eqref{EqnHanaSmileOne} implies that
$\big|(u + v)^T Q (u + v) \big| \leq 4 \UNICON_1 \Width(\USETPLUS) + 4
\delta$ for all $v \in \VSETPLUS$.  Thus, when
inequalities~\eqref{EqnHanaSmileOne} through~\eqref{EqnHanaSmileThree}
hold, the decomposition~\eqref{EqnCorrectDecomposition} implies that
\begin{align}
\label{EqnHanaBurp}
|u^T Q u| & \leq \frac{\UNICON_1}{2} \big \{ 4 \Width(\USETPLUS) +
\Width(\infset) + \Width(\{u\}) \big \} + 3 \delta.
\end{align}
It remains to simplify the sum of the three Gaussian complexity terms.
An easy calculation gives $\Width(\{u\}) \leq \sqrt{2/\pi} \leq
\Width(\infset)$.  In addition, we claim that
\begin{align}
\label{EqnDecomposeBound}
\Width(\VSET) & \leq \Width(\{u\}) + \Width(\infset).
\end{align}
Given any $v \in \VSETPLUS$, let $\Pi(v)$ denote its projection onto
the subspace orthogonal to $u$.  We can then write $v = \alpha u +
\Pi(v)$ for some scalar $\alpha \in [0,1]$, where $\|\Pi(v)\|_2 =
\sqrt{1 - \alpha^2}$.  In terms of this decomposition, we have
\begin{align*}
\|u + v\|^2_2 & = \|(1+ \alpha) u + \Pi(v) \|_2^2 \; = \; (1 +
\alpha)^2 + 1- \alpha^2 \; = \; 2 + 2 \alpha.
\end{align*}
Consequently, we have
\begin{align}
\Big |\inprod{g}{ \frac{u + v}{\|u + v\|_2}} \Big| & = \Big| \frac{(1
  + \alpha)}{\sqrt{2 (1 + \alpha)} } \inprod{g}{u} + \frac{1}{\sqrt{2
    (1 + \alpha)}} \inprod{g}{\Pi(v)} \Big| \nonumber \\
\label{EqnJelani}
& \leq \big|\inprod{g}{u}\big| + \big| \inprod{g}{\Pi(v)} \big|.
\end{align}
For any pair $v, v' \in \VSETPLUS$, note that
\begin{align*}
\var \big( \inprod{g}{\Pi(v)} - \inprod{g}{\Pi(v')} \big) & = \|\Pi(v)
- \Pi(v'\|_2^2 \leq \|v - v'\|_2^2 \; = \; \var \big( \inprod{g}{v} -
\inprod{g}{v'} \big).
\end{align*}
where the inequality follows by the non-expansiveness of projection.
Consequently, by the Sudakov-Fernique comparison, we have
\begin{align*}
\Exs \big[ \sup_{v \in \VSETPLUS} |\inprod{g}{\Pi(v)}| \big] & \leq
\Exs \big[ \sup_{v \in \VSETPLUS} |\inprod{g}{v}| \big] \; = \;
\Width(\VSETPLUS).
\end{align*}
Since $\VSETPLUS \subseteq \infset$, we have $\Width(\VSETPLUS) \leq
\Width(\infset)$.  Combined with our earlier inequality~\eqref{EqnJelani},
we have shown that
\begin{align*}
\Width(\USETPLUS) & \leq \Width(\{u\}) + \Width(\infset) \leq 2
\, \Width(\infset).
\end{align*}
Substituting back into our original upper bound~\eqref{EqnHanaBurp},
we have established that
\begin{align}
\label{EqnTechno}
\sup_{v \in \VSETPLUS} \big| u^T Q v \big| & \leq \frac{\UNICON_1}{2
  \sqrt{\numproj}} \big\{ 8 \Width(\infset) + 2 \Width(\infset) \big
\} + 3 \delta \; = \; \frac{5 \, \UNICON_1}{\sqrt{\numproj}} \,
\Width(\infset) + 3 \delta.
\end{align}
with high probability.

As for the supremum over $\VSETMINUS$, in this case, we use the
decomposition
\begin{align*}
u^T Q v & = \frac{1}{2} \Big \{ v^T Q v + u^T Q u - (v- u)^T Q (v - u)
\Big \}.
\end{align*}
The analogue of $\USETPLUS$ is the set $\USETMINUS = \{ \frac{v -
  u}{\|v - u\|_2} \, \mid \, v \in \VSETMINUS \}$.  Since
$\inprod{-u}{v} \geq 0$ for all $v \in \VSETMINUS$, the same argument
as before can be applied to show that $\sup_{v \in \VSETMINUS} |u^T Q v|$
satisfies the same bound~\eqref{EqnTechno} with high probability.

Putting together the pieces, we have established that, with
probability at least $1 - 6 \CEXP{- \UNICON_2 \numproj
  \delta^2/\sigma^4}$, we have
\begin{align*}
\ZSUP = \sup_{v \in \infset} \big| u^T Q v \big| & \leq \frac{10
  \UNICON_1}{\sqrt{\numproj}} \, \Width(\infset) + 6 \delta \;
\stackrel{(i)}{\leq} 9 \delta,
\end{align*}
where inequality (i) makes use of the assumed lower bound on the
projection dimension.  The claim follows by rescaling $\delta$ and
redefining the universal constants appropriately.


\subsection{Proof of Theorem~\ref{ThmHadamard}}
\label{SecProofHadmard}

We begin by stating two technical lemmas that provide control on the
random variables $\ZINF$ and $\ZSUP$ for randomized orthogonal
systems.  These results involve the $\Sketch$-Gaussian width
previously defined in equation~\eqref{EqnDefnSketchWidth}; we also
recall the Rademacher width
\begin{align}
\WidthRad(\Amat \ConeSet) & \defn \Exs_{\rade{}} \sup_{z \in \infset}
|\inprod{z}{\rade{}}|.
 \end{align}

\blems[Lower bound on $\ZINF$]
\label{LemROSZINF}
Given a projection size $\numproj$ satisfying the
bound~\eqref{EqnKeyLower} for a sufficiently large universal constant
$\UNICON_0$, we have
\begin{align}
\underbrace{\inf_{v \in \Amat \ConeSet \cap \SPHERE{\numobs}}
  \frac{1}{\numproj} \| \Sketch v\|^2_2}_{\ZINF} & \geq 1 - \MYLEMEPS
\end{align}
with probability at least $\STRANGEPROB$.
\elems

\blems[Upper bound on $\ZSUP$]
\label{LemROSZSUP}
Given a projection size $\numproj$ satisfying the
bound~\eqref{EqnKeyLower} for a sufficiently large universal constant
$\UNICON_0$, we have
\begin{align}
\underbrace{\sup_{v \in \Amat \ConeSet \cap \SPHERE{\numobs}} \Big|
  \inprod{\fixvec}{(\frac{\Sketch^T \Sketch}{m} - I) \, v} \Big|}_{\ZSUP} & \leq
\MYLEMEPS
\end{align}
with probability at least $\STRANGEPROB$.
\elems

Taking them as given, the proof of Theorem~\ref{ThmHadamard} is easily
completed.  Based on a combination of the two lemmas, for any $\delta
\in [0, 1/2]$, we have
\begin{align*}
2 \frac{\ZSUP}{\ZINF} & \leq \frac{2 \delta}{1 - \delta} \; \leq \; 4
\delta,
\end{align*}
with probability at least $\STRANGEPROB$.  The claimed form of the
bound follows via the rescaling $\delta \mapsto 4 \delta$, and
suitable adjustments of the universal constants. \\

\noindent In the following, we use $\Ball_2^\numobs = \{z \in
\real^\numobs \, \mid \, \|z\|_2 \leq 1 \}$ to denote the Euclidean
ball of radius one in $\real^\numobs$.
\bprops
\label{PropROSRad} 
Let $\{\sketch_i\}_{i=1}^\numproj$ be i.i.d. samples from a randomized
orthogonal system.  Then for any subset $\YSET \subseteq
\Ball_2^\numobs$ and any $\delta \in [0,1]$ and $\HACKKAP > 0$, we
have
\begin{align}
\sup_{y \in \YSET} \Big| y^T \Big( \frac{\Sketch^T \Sketch}{\numproj}
- I \Big) y \big| & \leq 8 \Big \{ \WidthRad(\YSET) + \sqrt{2 (1 +
  \HACKKAP) \, \log (\numproj \numobs)} \Big \} \:
\frac{\PLSKETCHWIDTH(\YSET)}{\sqrt{\numproj}} + \frac{\delta}{2}
\end{align}
with probability at least $\STRANGEPROBY$.
\eprops
%


\subsubsection{Proof of Lemma~\ref{LemROSZINF}}

This lemma is an immediate consequence of Proposition~\ref{PropROSRad}
with $\YSET = \Amat \ConeSet \cap \SPHERE{\numobs}$ and $\HACKKAP =
2$. In particular, with a sufficiently large constant $\UNICON_0$, the
lower bound~\eqref{EqnKeyLower} on the projection dimension ensures
that $8 \Big \{ \WidthRad(\YSET) + \sqrt{6 \, \log (\numproj \numobs)}
\Big \} \leq \frac{\delta}{2}$, from which the claim follows.

\subsubsection{Proof of Lemma~\ref{LemROSZSUP}}

We again introduce the convenient shorthand $Q = \frac{\Sketch^T
  \Sketch}{\numproj} - I$.  For any subset $\YSET \subseteq
\Ball_2^\numobs$, define the random variable $Z_0(\YSET) = \sup_{y \in
  \YSET} |y^T Q y|$.  Note that Proposition~\ref{PropROSRad}
provides control on any such random variable.  Now given the fixed
unit-norm vector $u \in \real^\numobs$, define the set 
\begin{align*}
\VSET = \frac{1}{2} \{u + v \, \mid \, v \in \infset \}.
\end{align*}
Since $\|u + v\|_2 \leq \|u\|_2 + \|v\|_2 = 2$, we have the inclusion
$\VSET \subseteq \Ball_2^\numobs$.  For any $v \in \infset$, the
triangle inequality implies that
\begin{align*}
\big|u^T Q v \big| & = 4 \big | \big(\frac{u + v}{2} \big)^T Q \frac{u
  + v}{2} \big| + \big| v^T Q v \big| + \big| u^T Q u \big| \\
& \leq 4 Z_0(\VSET) + Z_0(\infset) + Z_0(\{u\}).
\end{align*}
We now apply Proposition~\ref{PropROSRad} in three times in succession
with the sets $\YSET = \VSET$, $\YSET = \infset$ and $\YSET = \{u\}$,
thereby finding that
\begin{align*}
\big|u^T Q v \big| & \leq \frac{1}{\sqrt{\numproj}} \Big \{ 4
\Phi(\VSET) + \Phi(\infset) + \Phi(\{u\}) \Big \} + 3 \delta,
\end{align*}
where we have defined the set-based function
\begin{align*}
\Phi(\YSET) & = 8 \Big \{ \WidthRad(\YSET) + \sqrt{6 \, \log (\numproj
  \numobs)} \Big \} \: \PLSKETCHWIDTH(\YSET)
\end{align*}
By inspection, we have $\WidthRad(\{u\}) \leq 1 \leq 2
\WidthRad(\infset)$ and $\PLSKETCHWIDTH(\{u\}) \leq 1 \leq 2
\SKETCHWIDTH$, and hence $\Phi(\{u\}) \leq 2 \Phi(\infset)$.
Moreover, by the triangle inequality, we have
\begin{align*}
\WidthRad(\VSET) & \leq \Exs_{\rade{}} |\inprod{\rade{}}{u}| +
\Exs_{\rade{}} \big[\sup_{v \in \infset} |\inprod{\rade{}}{v}| \; \leq
  1 + \WidthRad(\infset) \; \leq 4 \WidthRad(\infset).
\end{align*}
A similar argument yields $\PLSKETCHWIDTH(\VSET) \leq 3 \SKETCHWIDTH$,
and putting together the pieces yields
\begin{align*}
\Phi(\VSET) & \leq 8 \big \{ 3 \WidthRad(\infset) + \sqrt{6
  \log(\numproj \numobs)} \big \} \, (3 \, \SKETCHWIDTH) \; \leq \; 9
\Phi(\infset).
\end{align*}
Putting together the pieces, we have shown that for any $v \in
\infset$, 
\begin{align*}
|u^T Q v | & \leq \frac{39}{\sqrt{\numproj}} \Phi(\infset) + 3 \delta.
\end{align*}
Using the lower bound~\eqref{EqnKeyLower} on the projection dimension,
we are have $\frac{39}{\sqrt{\numproj}} \Phi(\infset) \leq \delta$,
and hence $\ZSUP \leq 4 \delta$ with probability at least
$\STRANGEPROB$.  A rescaling of $\delta$, along with suitable
modification of the numerical constants, yields the claim.


\subsubsection{ Proof of Proposition~\ref{PropROSRad}}

We first fix on the diagonal matrix $\DD = \diag(\nu)$, and compute
probabilities over the randomness in the vectors $\sketchtil_i =
\sqrt{\numobs} H^T p_i$, where the picking vector $p_i$ is chosen
uniformly at random.  Using $\mprob_P$ to denote probability taken
over these i.i.d. choices, a symmetrization argument~ (see \cite{Pollard84}, p. 14) 
yields
\begin{align*}
\mprob_P\big[ \ZZERO \geq t] & \leq 4 \, \mprob_{\rade{}, P} \Big[
  \underbrace{\sup_{z \in \infset} \big| \frac{1}{\numproj}
    \sum_{i=1}^\numproj \rade{i} \inprod{\sketchtil_i}{D z}^2
    \big|}_{\ZZERO'} \geq \frac{t}{4} \Big],
\end{align*}
where $\{\rade{i}\}_{i=1}^\numproj$ is an i.i.d. sequence of
Rademacher variables.  Now define the function \mbox{$g: \{-1,
  1\}^\usedim \rightarrow \real$} via
\begin{align}
\label{EqnDefnG}
g(\nu) & \defn \Exs_{\rade{},P} \Big[ \sup_{y \in \YSET} \big|
  \frac{1}{\numproj} \sum_{i=1}^\numproj \rade{i}
  \inprod{\sketchtil_i}{\diag(\nu) y} \big| \big].
\end{align}
Note that $\Exs [g(\nu)] = \PLSKETCHWIDTH(\YSET)$ by construction.
For a truncation level $\tau > 0$ to be chosen, define the events
\begin{align}
\PLGOODONE \defn \big \{ \max_{j = 1, \ldots, \numobs} \sup_{y \in
  \YSET} |\inprod{\sqrt{\numobs} h_j}{\diag(\nu) y}| \leq \tau \big
\}, \quad \mbox{and} \quad \PLGOODTWO \defn \big\{ g(\nu) \leq
\PLSKETCHWIDTH(\YSET) + \frac{\delta}{32 \trunlev} \big \}.
\end{align}
To be clear, the only randomness involved in either event is over the
Rademacher vector \mbox{$\nu \in \{-1, +1 \}^\numobs$.}  We then
condition on the event $\PLGOOD = \PLGOODONE \cap \PLGOODTWO$ and its
complement to obtain
\begin{align*}
\mprob_{\rade{}, P, \nu} \big[ \ZZERO^\prime \geq t \big] & = \Exs
\Big \{ \Ind[\ZZERO^\prime \geq t] \, \Ind[\PLGOOD] +
\Ind[\ZZERO^\prime \geq t] \Ind[\PLGOOD^c] \Big \}
\; \leq \; \mprob_{\rade{}, P} \big[ \ZZERO^\prime \geq t \mid \nu \in
  \PLGOOD \big] \; \mprob_\nu[\PLGOOD] + \mprob_\nu[\PLGOOD^c].
\end{align*}
We bound each of these two terms in turn.

\blems
\label{LemMainTerm}
For any $\delta \in [0,1]$, we have
\begin{align}
\label{EqnCamus}
\mprob_{\rade{}, P} \big[ \ZZERO^\prime \geq 2 \tau
  \PLSKETCHWIDTH(\YSET) + \frac{\delta}{8} \mid \PLGOOD] \;
\mprob_D[\PLGOOD] & \leq \UNICON_1 \CEXP{ - \UNICON_2 \frac{\numproj
    \delta^2}{\tau^2}}.
\end{align}
\elems

\blems
\label{LemPLGOOD}
With truncation level $\trunlev = \MYRAD(\YSET) + \sqrt{2 (1 +
  \HACKKAP) \, \log (\numproj \numobs)}$ for some $\HACKKAP > 0$, we have
\begin{align}
\mprob_\nu[\PLGOOD^c] & \leq \frac{1}{(\numproj \numobs)^\HACKKAP} +
\CEXP{-\frac{\numproj \delta^2}{4 \trunlev^2}}.
\end{align}
\elems

\noindent See Appendix~\ref{AppPLGOOD} for the proof of these two
claims. \\

\noindent Combining Lemmas~\ref{LemMainTerm} and~\ref{LemPLGOOD}, we
conclude that
\begin{align*}
\mprob_{P, \nu}[Z \geq 8 \tau \PLSKETCHWIDTH(\YSET) +
  \frac{\delta}{2}] & \leq 4 \mprob_{\rade{}, P, \nu}[\ZZERO^\prime
  \geq 2 \tau \PLSKETCHWIDTH(\YSET) + \frac{\delta}{8}] \; \leq \; c_1
\CEXP{- c_2 \frac{\numproj \delta^2}{\tau^2}} + \frac{1}{(\numproj
  \numobs)^\HACKKAP},
\end{align*}
as claimed.  


\section{Techniques for sharpening bounds}
\label{SecSharpen}

In this section, we provide some technique for obtaining sharper
bounds for randomized orthonormal systems when the underlying tangent
cone has particular structure.  In particular, this technique can be
used to obtain sharper bounds for subspaces, $\ell_1$-induced cones,
as well as nuclear norm cones.


\subsection{Sharpening bounds for a subspace}
\label{SecSubspaceSharpen}

As a warm-up, we begin by showing how to obtain sharper bounds when
$\ConeSet$ is a subspace.  For instance, this allows us to obtain the
result stated in Corollary~\ref{CorUncLS}(b).  Consider the random
variable
\begin{align*}
Z(\Amat \ConeSet) & = \sup_{z \in \infset} \big| z^T Q z \big|, \quad
\mbox{where $Q = \frac{\Sketch^T \Sketch}{\numproj} - I$.}
\end{align*}
For a parameter $\epsilon \in (0,1)$ to be chosen, let $\{z^1, \ldots,
z^M\}$ be an $\epsilon$-cover of the set $\infset$.  For any $z \in
\infset$, there is some $j \in [M]$ such that $z = z^j + \Delta$,
where $\|\Delta\|_2 \leq \epsilon$.  Consequently, we can write
\begin{align*}
\big| z^T Q z \big| & \leq |(z^j)^T Q z^j| + 2 |\Delta^T Q z^j| +
|\Delta^T Q \Delta|
\end{align*}
Since $\Amat \ConeSet$ is a subspace, the difference vector $\Delta$
also belongs to $\Amat \ConeSet$.  Consequently, we have $|\Delta^T Q
z^j| \leq \epsilon Z(\Amat \ConeSet)$ and $|\Delta^T Q \Delta| \leq
\epsilon^2 Z(\Amat \ConeSet)$.  Putting together the pieces, we have
shown that
\begin{align*}
(1 - 2 \epsilon - \epsilon^2) Z(\Amat \ConeSet) & \leq \max_{j = 1,
    \ldots, M} |(z^j)^T Q z^j|.
\end{align*}
Setting $\epsilon = 1/8$ yields that $Z(\Amat \ConeSet) \leq
\frac{3}{2} \max_{j = 1, \ldots, M} |(z^j)^T R z^j|$.

Having reduced the problem to a finite maximum, we can now make use of
JL-embedding property of a randomized orthogonal system proven in
Theorem 3.1 of Krahmer and Ward~\cite{KraWar11}: in particular, their
theorem implies that for any collection of $M$ fixed points $\{z^1,
\ldots, z^M \}$ and $\delta \in (0,1)$, a ROS sketching matrix $S \in
\real^{\numproj \times \numobs}$ satisfies the bounds
\begin{align}
(1-\delta)\|z^j\|_2^2 \le \frac{1}{m}\|S z^j\|_2^2 &\le (1+\delta)\|z^j\|_2^2
  \qquad \mbox{for all $j = 1, \ldots, M$}
\end{align}
with probability $1-\eta$ if $\numproj \geq \frac{c}{\delta^2}
\log^4(\numobs ) \log(\frac{M}{\eta})$.  For our chosen collection, we
have $\|z^j\|_2 = 1$ for all $j = 1, \ldots, M$, so that our
discretization plus this bound implies that $Z(\Amat \ConeSet) \leq
\frac{3}{2}\delta$.  Setting $\eta = \CEXP{-c_2 \numproj \delta^2}$
for a sufficiently small constant $c_2$ yields that this bound holds
with probability $1 - \CEXP{-c_2 \numproj \delta^2}$.

The only remaining step is to relate $\log M$ to the Gaussian width of
the set.  By the Sudakov minoration~\cite{LedTal91} and recalling that
$\epsilon = 1/8$, there is a universal constant $c > 0$ such that
\begin{align*}
\sqrt{\log M} & \leq c \; \Width(\Amat \ConeSet) \;
\stackrel{(i)}{\leq} c \; \sqrt{\rank(\Amat)},
\end{align*}
where the final inequality (i) follows from our previous
calculation~\eqref{EqnSubspaceWidth} in the proof of
Corollary~\ref{CorUncLS}.


\subsection{Reduction to finite maximum}

The preceding argument suggests a general scheme for obtaining sharper
results, namely by reducing to finite maxima.  In this section, we
provide a more general form of this scheme.  It applies to random
variables of the form
\begin{align}
Z(\YSET) & = \sup_{y \in \YSET} \big| y^T \big( \frac{\Amat^T
  \Sketch^T \Sketch \Amat}{\numproj} - I \big) y \big|, \qquad
\mbox{where $\YSET \subset \real^\usedim$.}
\end{align}
For any set $\YSET$, we define the first and second set differences as
\begin{align}
\SetDiff{\YSET} & \defn \YSET- \YSET \; = \big \{ y - y' \, \mid y, y'
\in \YSET \big \} \quad \mbox{and} \quad \SetDiffTwo{\YSET} \, \defn
\, \SetDiff{\SetDiff{\YSET}}.
\end{align}
Note that $\YSET \subseteq \SetDiff{\YSET}$ whenever $0 \in \YSET$.
Let $\SphereProj{\YSET}$ denote the projection of $\YSET$ onto the
Euclidean sphere $\SPHERE{\usedim}$.  \\

\noindent With this notation, the following lemma shows how to reduce
bounding $Z(\YSET_1)$ to taking a finite maximum over a cover of
$\YSET_0$:\\
\blems
\label{LemmaInclusion} 
Consider a pair of sets $\YSET_0$ and $\YSET_1$ such that $0 \in
\YSET_0$, the set $\YSET_1$ is convex, and for some constant $\alpha
\geq 1$, we have
\begin{align}
\label{EqnMertInclusions}
(a) \; \YSET_1 \subseteq \; \clconv (\YSET_0), \quad (b) \;
  \SetDiffTwo{\YSET_0} \subseteq \alpha \YSET_1, \quad \mbox{and} \quad
  (c) \; \SphereProj{\SetDiffTwo{\YSET_0}} \subseteq \alpha \YSET_1.
\end{align}
Let $\{z^1, \ldots,z^M \}$ be an $\epsilon$-covering of the set
$\SetDiff{\YSET_0}$ in Euclidean norm for some $\epsilon \in (0,
\frac{1}{27 \alpha^2}]$.  Then for any symmetric matrix $Q$, we have
\begin{align}
\label{EqnMertDiscretize}
\sup_{z \in \YSET_1} |z^T Q z| & \leq 3 \max_{j = 1, \ldots, M}
|(z^j)^T Q z^j|.
\end{align}
\elems 
\noindent See Appendix~\ref{AppSharpen} for the proof of this lemma.
In the following subsections, we demonstrate how this auxiliary result
can be used to obtain sharper results for various special cases.


\subsection{Sharpening $\ell_1$-based bounds}
\label{SecEllOneSharpen}

The sharpened bounds in Corollary~\ref{CorLasso} are based on the
following lemma.  It applies to the tangent cone $\ConeSet$ of the
$\ell_1$-norm at a vector $\xstar$ with $\ell_0$-norm equal to
$\kdim$, as defined in equation~\eqref{EqnEllOneTangentCone}.

\blems
\label{LemSharpEllone}
For any $\delta \in (0,1)$, a projection dimension lower bounded as
$\numproj \geq \frac{c_0}{\delta^2} \,
\big(\frac{\sigkmaxsq(\Amat)}{\sigkminsq(\Amat)}\big)^2 \, \kdim
\log^5(\usedim)$ guarantees that
\begin{align}
\label{EqnDiscreteInfVecFinal}
\sup_{v \in \infset } |v (\frac{\Sketch^T \Sketch}{m} - I) v| & \leq \delta
\end{align}
with probability at least $\ROSHACK$.
\elems
\HACKPROOF 
Any $v \in \infset$ has the form $v = \Amat u$ for some $u \in
\ConeSet$.  Any $u \in \ConeSet$ satisfies the inequality $\|u\|_1
\leq 2 \sqrt{\kdim} \|u\|_2$, so that by definition of the
$\ell_1$-restricted eigenvalue~\eqref{EqnDefnRE}, we are guaranteed
that $\sigkminsq(\Amat) \|u\|^2_2 \leq \|A u \|^2_2 = 1$.  Putting
together the pieces, we conclude that
\begin{align*}
\sup_{v \in \infset } |v (\Sketch^T \Sketch - I) v| & \leq
\frac{1}{\sigkminsq(\Amat)} \sup_{y \in \YSET_1} \Big| y
\big(\frac{\Amat^T \Sketch^T \Sketch \Amat}{\numproj} - \Amat^T \Amat
\big) y \Big| \; = \; \frac{1}{\sigkminsq(\Amat)} \, Z(\YSET_1),
\end{align*}
where
\begin{align*}
\YSET_1 & = \Ball_2(1) \cap \Ball_1(2\sqrt{\kdim}) \; = \; \big \{
\Delta \in \real^{\matdim} \mid \|\Delta\|_1 \leq 2\sqrt{\kdim}, \;
\|\Delta\|_2\leq 1 \big \}.
\end{align*}
Now consider the set
\begin{align*}
\YSET_0 & = \; \Ball_2(3) \cap \Ball_0(4\kdim) \; = \; \big \{ \;
\Delta \in \real^{\usedim} \mid \|\Delta\|_{0} \leq 4\kdim, \;
\|\Delta\|_2 \leq 3 \big \},
\end{align*}
We claim that the pair $(\YSET_0, \YSET_1)$ satisfy the conditions of
Lemma~\ref{LemmaInclusion} with $\alpha = 24$. The
inclusion~\eqref{EqnMertInclusions}(a) follows from Lemma 11 in the
paper~\cite{LohWai11}; it is also a consequence of a more general
result to be stated in the sequel as Lemma~\ref{LemSandwich}.  Turning
to the inclusion~\eqref{EqnMertInclusions}(b), any vector $v \in
\SetDiffTwo{\YSET_0}$ can be written as $y - y' - (x - x')$ with $x,
x', y, y' \in \YSET_0$, whence $\|v\|_0 \leq 16 \kdim$ and $\|v\|_2
\leq 12$.  Consequently, we have $\|v\|_1 \leq 4 \sqrt{\kdim}
\|v\|_2$.  Rescaling by $1/12$ shows that $\SetDiffTwo{\YSET_0}
\subseteq 24 \YSET_1$.  A similar argument shows that
$\SphereProj{\SetDiffTwo{\YSET_0}}$ satisfies the same containment.

Consequently, applying Lemma~\ref{LemmaInclusion} with the symmetric
matrix $R = \frac{\Amat^T \Sketch^T \Sketch \Amat}{\numproj} - \Amat^T
\Amat$ implies that
\begin{align*}
Z(\YSET_1) & \leq 3 \max_{j = 1, \ldots, M} |(z^j)^T R z^j|,
\end{align*}
where $\{z^1, \ldots, z^M\}$ is an $\frac{1}{27 \alpha^2}$ covering of
the set $\YSET_0$. By the JL-embedding result of Krahmer and
Ward~\cite{KraWar11}, taking $\numproj > \frac{c}{\delta^2} \log^4
\usedim \, \log(M/\eta)$ samples suffices to ensure that, with
probability at least $1 - \eta$, we have
\begin{align}
\label{EqnHot}
\max_{j = 1, \ldots, M} |(z^j)^T R z^j| & \leq \delta \, \max_{j = 1,
  \ldots, M} \|\Amat z^j\|_2^2.
\end{align}
By the Sudakov minoration~\cite{LedTal91} and recalling that $\epsilon
= \frac{1}{27 \alpha^2}$ is a fixed quantity, we have
\begin{align}
\sqrt{\log M} & \leq c' \, \Width(\YSET_0) \; \leq \; c'' \sqrt{\kdim
  \log \usedim}, \label{EqnSharpElloneSud}
\end{align} 
where the final step follows by an easy calculation.  Since $\|z^j\|_2
= 1$ for all $j \in [M]$, we are guaranteed that \mbox{$\max_{j = 1,
    \ldots, M} \|\Amat z^j\|_2^2 \leq \sigkmaxsq(\Amat)$,} so that our
earlier bound~\eqref{EqnHot} implies that as long as $\numproj >
\frac{c}{\delta^2} \kdim \log (\usedim) \log^4 \numobs$, we have
\begin{align*}
\sup_{v \in \infset } |v (\frac{\Sketch^T \Sketch}{m} - I) v| & \leq 3 \delta
\,\frac{\sigkmaxsq(\Amat)}{\sigkminsq(\Amat)}
\end{align*}
with high probability.  Applying the rescaling $\delta \mapsto
\frac{\sigkminsq(\Amat)}{\sigkmaxsq(\Amat)} \delta$ yields the claim.
\HACKENDPROOF

\blems 
\label{LemSharpElloneAnnoying} Let $\uvec \in \SPHERE{\usedim}$
be a fixed vector.  Under the conditions of
Lemma~\ref{LemSharpEllone}, we have
\begin{align}
\label{EqnMertBound}
\max_{v \in \Amat \ConeSet \cap \SPHERE{\numobs}} \big |u (\frac{\Sketch^T \Sketch}{m} - I)
v \big| & \leq \delta
\end{align}
with probability at least $\ROSHACKN$.  
\elems

\HACKPROOF 
Throughout this proof, we make use of the convenient shorthand $Q =
\frac{\Sketch^T \Sketch}{\numproj} - I$.  Choose the sets $\YSET_0$
and $\YSET_1$ as in Lemma~\ref{LemSharpEllone}.  Any $v \in \infset$
can be written as $v = A z$ for some $z \in \ConeSet$, and for which
$\|z\|_2 \leq \frac{\|A z\|_2}{\sigkminsq(\Amat)}$.  Consequently,
using the definitions of $\YSET_0$ and $\YSET_1$, we have
\begin{align}
\max_{v \in \infset} |u^T Q v| \; \leq \; \frac{1}{\sigkminsq(\Amat)}
\max_{z \in \YSET_1} \big| u^T Q A z \big| & \leq
\frac{1}{\sigkminsq(\Amat)} \max_{z \in \clconv(\YSET_0)} \big|u^T Q A
z \big| \nonumber \\
\label{EqnLemSelfBoundingAnnoy}
& = \frac{1}{\sigkminsq(\Amat)} \max_{z \in \YSET_0} \big|u^T Q A z
\big|,
\end{align}
where the last equality follows since the supremum is attained at an
extreme point of $\YSET_0$. 

For a parameter $\epsilon \in (0,1)$ to be chosen, let $\{z^1,
\ldots,z^M \}$ be a $\epsilon$-covering of the set $\YSET_0$ in the
Euclidean norm.  Using this covering, we can write
\begin{align*}
 \sup_{z \in \YSET_0} \big|u^T Q A z \big| & \leq \max_{j \in [M]}
 \big|u^T Q A z^j \big| + \sup_{\Delta \in \SetDiff{\YSET_0}, \;
   \|\Delta\|_2 \leq \epsilon} \big| u^T Q A \Delta \big | \\
& = \max_{j \in [M]} \big|u^T Q A z^j \big| + \epsilon \sup_{\Delta
     \in \Pi(\SetDiff{\YSET_0})} \big|u^T Q A \Delta \big| \\
& \leq \max_{j \in [M]} \big|u^T Q A z^j \big| + \epsilon \alpha
 \sup_{\Delta \in \YSET_1} \big|u^T Q A \Delta \big|.
\end{align*}
Combined with equation~\eqref{EqnLemSelfBoundingAnnoy}, we conclude
that
\begin{align}
\label{EqnLemSelfBoundingAnnoyDiscrete}
\sup_{z \in \YSET_1} \big|u^T Q A z \big| \le \frac{1}{1-\epsilon
  \alpha} \max_{j \in [M]} \big|u^T Q A z^j \big|.
\end{align}
For each $j \in [M]$, we have the upper bound
\begin{align}
\big|u^T Q A z^j \big| & \leq |(A{z^j}+u)^T Q (A {z^j} + u)| + |(A
z^j)^T Q A{z^j}| + |u^T Q u|.
\end{align}
Based on this decomposition, we apply the JL-embedding
property~\cite{KraWar11} to ROS matrices to the collection of $2M+1$
points given by $\cup_{j\in [M]}\{ Az^j, Az^j+ u,\} \cup \{u\}$.
Doing so ensures that, for any fixed $\delta \in (0,1)$, we have
\begin{align*}
\max_{j\in [M]} \big|u^T Q A z^j \big| & \leq \delta \big(
\|Az^{j}+u\|_2^2 + \|Az^{j}\|_2^2 + \|u\|_2^2 \big).
\end{align*}
with probability $1-\eta$ as long as $\numproj > \frac{c_0}{\delta^2}
\log^4(\numobs) \, \log \big( \frac{2M+1}{\eta} \big)$.  Now observe
that
\begin{align*}
\|Az^{j} + u\|_2^2 + \|A z^j\|_2^2 + \|u\|_2^2 & \leq 3 \|A z^j\|_2^2
+ 3 \|u\|_2^2 \, \leq 3 \big( \sigkmaxsq(\Amat) + 1 \big),
\end{align*}
and consequently, we have $\max_{j\in [M]} \big|u^T Q A z^j \big| \leq
3 \delta \, \Big( {\sigkmaxsq(\Amat)} + 1 \Big)$.  Setting $\epsilon =
\frac{1}{2\alpha}$, $\eta = e^{-c_1 \frac{\numproj
    \delta^2}{\log^4(\numobs)}}$ and combining with our earlier
bound~\eqref{EqnLemSelfBoundingAnnoyDiscrete}, we conclude that
\begin{align}
\sup_{v \in \infset} |u^T (\frac{\Sketch^T \Sketch}{\numproj} - I) A
v| & \leq 6 \delta \, \frac{\big(\sigkmaxsq(\Amat) + 1
  \big)}{\sigkminsq(\Amat) }
\end{align}
with probability $\ROSHACKN$.  Combined with the covering number
estimate from equation~\eqref{EqnSharpElloneSud}, the claim follows.

\HACKENDPROOF


\subsection{Sharpening nuclear norm bounds}
\label{SecNuclearSharpen}

We now show how the same approach may also be used to derive sharper
bounds on the projection dimension for nuclear norm regularization.
As shown in Lemma 1 in the paper~\cite{NegWai09}, for the nuclear norm
ball $\nucnorm{X} \leq R$, the tangent cone at any rank $\rdim$ matrix
is contained within the set
\begin{align}
\label{EqnNucNormCone}
\ConeSet & \defn \big \{ \Delta \in \real^{\usedim_1 \times \usedim_2}
\, \mid \, \nucnorm{\Delta} \leq 2 \sqrt{\rdim} \frobnorm{\Delta} \big
\},
\end{align}
and accordingly, our analysis focuses on the set $\Aop \ConeSet \cap
\Sphere{\numobs}$, where $\Aop: \real^{\usedima \times \usedimb}
\rightarrow \real^\numobs$ is a general linear operator.

In analogy with the sparse restricted eigenvalues~\eqref{EqnDefnRE},
we define the rank-constrained eigenvalues of the general operator
$\Aop: \real^{\usedima \times \usedimb} \rightarrow \real^\numobs$ as
follows:
\begin{align}
\label{EqnDefnRENuclear}
\sigrminsq(\Aop) \defn \min_{ \substack{\frobnorm{Z}= 1 \\ \nucnorm{Z}
    \leq 2\sqrt{\rdim}}} \|\Aop(Z)\|^2_2, \quad \mbox{and} \quad
\sigrmaxsq(\Aop) \defn \max_{ \substack{\frobnorm{Z} = 1
    \\ \nucnorm{Z} \le 2\sqrt{\rdim}}} \|\Aop(Z)\|^2_2.
\end{align}

\blems
\label{LemNuclearSharpen}
Suppose that the optimum $X^*$ has rank at most $\rdim$.  For any
$\delta \in (0,1)$, a ROS sketch dimension lower bounded as $\numproj
\geq \frac{c_0}{ \delta^2}
\big(\frac{\sigrmaxsq(\Aop)}{\sigrminsq(\Aop)} \big)^2 \, {\rdim
  (\usedim_1 + \usedim_2) \log^4(\usedima \usedimb)}$ ensures that
\begin{align}
\label{EqnDiscreteInfVecFinalNuclear}
\sup_{z \in \Aop\ConeSet \cap \Sphere{\numobs} } |z (\frac{\Sketch^T \Sketch}{m} - I) z|
\leq \delta
\end{align}
with probability at least $\ROSHACKMAT$.
\elems
\HACKPROOF 
For an integer $\rdim \geq 1$, consider the sets
\begin{subequations}
\begin{align}
\label{EqnMertOne}
\YSET_1(\rdim) & = \Ball_F(1) \cap \Ball_{nuc}(2\sqrt{\rad}) \; = \;
\big \{ \Delta \in \real^{\usedim_1 \times \usedim_2}\mid
\nucnorm{\Delta} \leq 2\sqrt{\rad}, \; \frobnorm{\Delta} \leq 1 \big
\}, \quad \mbox{and} \\
\label{EqnMertZero}
\YSET_0(\rdim) & = \; \big \{ \Ball_F(3) \cap \Ball_{rank}(4\rad) \big
\} \; = \; \; \big \{ \Delta \in \real^{n_1\times n_2}\mid
\matsnorm{\Delta}{0} \leq 4\rad, \; \frobnorm{\Delta} \leq 3 \big \}.
\end{align}
\end{subequations}
In order to apply Lemma~\ref{LemmaInclusion} with this pair, we must
first show that the inclusions~\eqref{EqnMertInclusions} hold.
Inclusions (b) and (c) hold with $\alpha = 12$, as in the preceding
proof of Lemma~\ref{LemSharpEllone}.  Moreover, inclusion (a) also
holds, but this is a non-trivial claim stated and proved separately as
Lemma~\ref{LemSandwich} in Appendix~\ref{AppLemSandwich}.

Consequently, an application of Lemma~\ref{LemmaInclusion} with the
symmetric matrix $Q = \frac{\Aop^* \Sketch^T \Sketch \Aop}{\numproj} -
\Aop^* \Aop$ in dimension $\usedima \usedimb$ guarantees that
\begin{align*}
Z(\YSET_1(\rdim)) & \leq 3 \max_{j = 1, \ldots, M} |(z^j)^T Q z^j|,
\end{align*}
where $\{z^1, \ldots, z^M\}$ is a $\frac{1}{27 \alpha^2}$-covering of
the set $\YSET_0(\rdim)$.  By arguing as in the preceding proof of
Lemma~\ref{LemSharpEllone}, the proof is then reduced to upper
bounding the Gaussian complexity of $\YSET_0(\rdim)$.  Letting $G \in
\real^{\usedima \times \usedimb}$ denote a matrix of i.i.d. $N(0,1)$
variates, we have
\begin{align*}
\Width(\YSET_0(\rdim)) \; = \; \Exs \big[ \sup_{\Delta \in
    \YSET_0(\rdim)} \tracer{G}{\Delta} \big] & \leq \: 6 \sqrt{\rdim}
\, \Exs[\opnorm{G}] \; \leq \, 6 \sqrt{\rdim} \, \big( \sqrt{\usedima}
+ \sqrt{\usedimb} \big),
\end{align*}
where the final line follows from standard results~\cite{DavSza01} on
the operator norms of Gaussian random matrices.
\HACKENDPROOF

\blems
Let $u \in \Sphere{\numobs}$ be a fixed vector. Under the assumptions
of Lemma~\ref{LemNuclearSharpen}, we have
\begin{align}
\sup_{z \in \Aop\ConeSet \cap \Sphere{\numobs} } |u (\frac{\Sketch^T \Sketch}{m} - I) z|
\leq \delta
\end{align}
with probability at least $\ROSHACKMAT$.
\elems
\noindent The proof parallels the proof of
Lemma~\ref{LemSharpElloneAnnoying}, and hence is omitted. Finally the
sharpened bounds follow from the above lemmas and the deterministic
bound~\eqref{EqnZ1Z2Ratio}.


\section{Discussion}
\label{SecDiscussion}

In this paper, we have analyzed random projection methods for
computing approximation solutions to convex programs.  Our theory
applies to any convex program based on a linear/quadratic objective
functions, and involving arbitrary convex constraint set.  Our main
results provide lower bounds on the projection dimension that suffice
to ensure that the optimal solution to sketched problem provides a
$\MYEPS$-approximation to the original problem.  In the sub-Gaussian
case, this projection dimension can be chosen proportional to the
square of the Gaussian width of the tangent cone, and in many cases,
the same results hold (up to logarithmic factors) for sketches based
on randomized orthogonal systems. This width depends both on the
geometry of the constraint set, and the associated structure of the
optimal solution to the original convex program.  We provided
numerical simulations to illustrate the corollaries of our theorems in
various concrete settings.

\subsection*{Acknowledgements}

Both authors were partially supported by Office of Naval Research MURI
grant N00014-11-1-0688, and National Science Foundation Grants
CIF-31712-23800 and DMS-1107000. In addition, MP was supported by a Microsoft Research Fellowship.

\appendix

\section{Technical details for Corollary~\ref{CorLasso}}
\label{AppCorROSLasso}

In this appendix, we show how the second term in the
bound~\eqref{EqnROSLasso} follows as a corollary of
Theorem~\ref{ThmHadamard}.  From our previous calculations in the
proof of Corollary~\ref{CorLasso}(a), we have
\begin{align}
\label{EqnRadWidth}
\WidthRad(\Amat \ConeSet) \; \leq \; \Exs_{\rade{}} \big[ \sup_{
    \substack{\|u\|_1 \leq 2 \sqrt{k}\|u\|_2 \\ \|A u\|_2 = 1}} \big|
  \inprod{u}{A^T \rade{}} \big| & \leq \frac{2
    \sqrt{\kdim}}{\sqrt{\sigkminsq(\Amat)}} \Exs[\|A^T
    \rade{}\|_\infty] \; \leq \; 6 \sqrt{\kdim \log \usedim} \max_{j
    =1, \ldots, \usedim} \frac{\|a_j\|_2}{\sqrt{\sigkminsq(\Amat)}}.
\end{align}
Turning to the $\Sketch$-Gaussian width, we have
\begin{align*}
\SKETCHWIDTH & = \Exs_{g, \Sketch} \Big[ \sup_{ \substack{\|u\|_1 \leq
      2 \sqrt{\kdim} \|u\|_2 \\ \|A u\|_2 = 1}} \Big|
  \inprod{g}{\frac{\Sketch \Amat u}{\sqrt{\numproj}}} \Big| \Big] \;
\leq \; \frac{2 \sqrt{\kdim}}{\sqrt{\sigkminsq(\Amat)}} \, \; \Exs_{g,
  \Sketch} \|\frac{\Amat^T \Sketch^T g}{\sqrt{\numproj}}\|_\infty.
\end{align*}
Now the vector $\Sketch^T g/\sqrt{\numproj}$ is zero-mean Gaussian with
covariance $\Sketch^T \Sketch/\numproj$.  Consequently
\begin{align*}
 \Exs_{g} \|\frac{\Amat^T \Sketch^T g}{\sqrt{\numproj}}\|_\infty &
 \leq 4 \max_{j = 1, \ldots \usedim} \frac{\|\Sketch
   a_j\|_2}{\sqrt{\numproj}} \, \sqrt{\log \usedim}.
\end{align*}
Define the event $\Event = \big \{ \frac{\|\Sketch
  a_j\|_2}{\sqrt{\numproj}} \leq 2 \|a_j\|_2 \quad \mbox{for $j = 1,
  \ldots, \usedim$} \big \}$.  By the JL embedding theorem of Krahmer
and Ward~\cite{KraWar11}, as long as $\numproj > c_0 \log^5(\numobs)
\log(\usedim)$, we can ensure that $\mprob[\Event^c] \leq
\frac{1}{\numobs}$.  Since we always have $\|\Sketch
a_j\|_2/\sqrt{\numproj} \leq \|a_j\|_2 \sqrt{\numobs}$, we can
condition on $\Event$ and its complement, thereby obtaining that
\begin{align*}
 \Exs_{g, S} \big[\|\frac{\Amat^T \Sketch^T
     g}{\sqrt{\numproj}}\|_\infty \big] & \leq 8 \max_{j = 1, \ldots
   \usedim} \|a_j\|_2 \, \sqrt{\log \usedim} + 4 \, \mprob[\Event^c]
 \; \sqrt{\numobs} \max_{j = 1, \ldots \usedim} \|a_j\|_2 \,
 \sqrt{\log \usedim} \\
& \leq 12 \max_{j = 1, \ldots
   \usedim} \|a_j\|_2 \, \sqrt{\log \usedim}.
\end{align*}
Combined with our earlier calculation, we conclude that
\begin{align*}
\SKETCHWIDTH & \leq \frac{\max \limits_{j = 1, \ldots, \usedim}
  \|a_j\|_2}{\sqrt{\sigkminsq(\Amat)}} \sqrt{\kdim \log \usedim}.
\end{align*}
Substituting this upper bound, along with our earlier upper bound on
the Rademacher width~\eqref{EqnRadWidth}, yields the claim as a
consequence of Theorem~\ref{ThmHadamard}.


\section{Technical lemmas for Proposition~\ref{PropROSRad}}
\label{AppPLGOOD}

In this appendix, we prove the two technical lemmas, namely
Lemma~\ref{LemMainTerm} and~\ref{LemPLGOOD}, that underlie the proof
of Proposition~\ref{PropROSRad}.


\subsection{Proof of Lemma~\ref{LemMainTerm}}
   
Fixing some $D = \diag(\nu) \in \PLGOOD$, we first bound the
deviations of $\ZZERO^\prime$ above its expectation using Talagrand's
theorem on empirical processes (e.g., see Massart~\cite{Massart00b}
for one version with reasonable constants). Define the random vector
$\sketchtil = \sqrt{\numobs} h$, where $h$ is a randomly selected row,
as well as the functions $g_y(\rade{}, \sketchtil) = \rade{}
\inprod{\sketchtil}{\diag(\nu) y}^2$, we have $\|g_z\|_\infty \leq
\tau^2$ for all $y \in \YSET$.  Letting $\sketchtil = \sqrt{\numobs}
h$ for a randomly chosen row $h$, we have
\begin{align*}
\var(g_y) & \leq \tau^2 \Exs[\inprod{\sketchtil}{\diag(\nu) y}^2] \; =
\; \tau^2,
\end{align*}
also uniformly over $y \in \YSET$.  Thus, for any $\nu \in \PLGOOD$,
Talagrand's theorem~\cite{Massart00b} implies that
\begin{align*}
\mprob_{\rade{}, P} \big[ \ZZERO^\prime \geq \Exs_{\rade{}, P}
  [\ZZERO^\prime] + \frac{\delta}{16}] & \leq \UNICON_1 \CEXP{ -
  \UNICON_2 \frac{\numproj \delta^2}{\tau^2}} \qquad \mbox{for all
  $\delta \in [0,1]$.}
\end{align*}

It remains to bound the expectation.  By the Ledoux-Talagrand
contraction for Rademacher processes~\cite{LedTal91}, for any $\nu \in
\PLGOOD$, we have
\begin{align*}
\Exs_{\rade{}, P}[\ZZERO^\prime] & \stackrel{(i)}{\leq} \; 2 \, \tau
\, \Exs_{\rade{}, P} \big[\sup_{y \in \YSET} \big| \frac{1}{\numproj}
  \sum_{i=1}^\numproj \rade{i} \inprod{\sketch_i}{z} \big| \big] \;
\stackrel{(ii)}{\leq} \; 2 \tau \big \{ \PLSKETCHWIDTH(\YSET) +
\frac{\delta}{32 \tau} \big \} \; = \; 2 \PLSKETCHWIDTH(\YSET) +
\frac{\delta}{16},
\end{align*}
where inequality (i) uses the inclusion $\nu \in \PLGOODONE$, and step
(ii) relies on the inclusion $\nu \in \PLGOODTWO$.  Putting together
the pieces yields the claim~\eqref{EqnCamus}.


\subsection{Proof of Lemma~\ref{LemPLGOOD}}

It suffices to show that
\begin{align*}
\mprob[\PLGOODONE^c] \leq \frac{1}{(\numproj \numobs)^\HACKKAP} \quad
\mbox{and} \quad \mprob[\PLGOODTWO^c] \leq c_1 \CEXP{-c_2 \numproj
  \delta^2}.
\end{align*}
We begin by bounding $\mprob[\PLGOODONE^c]$.  Recall $\sketch_i^T =
\sqrt{\numobs} p_i^T H \diag(\nu)$, where $\nu \in \{-1, +1\}^\numobs$
is a vector of i.i.d. Rademacher variables.  Consequently, we have
$\inprod{\sketch_i}{y} = \sum_{j=1}^\numobs (\sqrt{\numobs} H_{ij})
\nu_j y_j$.  Since $|\sqrt{\numobs} H_{ij}| = 1$ for all $(i,j)$, the
random variable $\inprod{\sketch_i}{y}$ is equal in distribution to
the random variable $\inprod{\nu}{y}$.  Consequently, we have the
equality in distribution
\begin{align*}
\sup_{y \in \YSET} \big| \inprod{\sqrt{\numobs } p_i^T H
  \diag(\nu)}{y} \big| & \stackrel{d}{=} \underbrace{\sup_{y \in
    \YSET} \big| \inprod{\nu}{y} \big|}_{f(\nu)}.
\end{align*}
Since this equality in distribution holds for each $i = 1, \ldots,
\numobs$, the union bound guarantees that
\begin{align*}
\mprob[\PLGOODONE^c] & \leq \numobs \; \mprob \big[f(\nu) > \trunlev
  \big].
\end{align*}
Accordingly, it suffices to obtain a tail bound on $f$.  By
inspection, the the function $f$ is convex in $\nu$, and moreover
$|f(\nu) - f(\nu')| \leq  \|\nu - \nu'\|_2$, so that it is $1$-Lipschitz.
Therefore, by standard concentration results~\cite{Ledoux01}, we have
\begin{align}
\mprob \big[f(\nu) \geq \Exs[f(\nu)] + t \big] & \leq
\CEXP{-\frac{t^2}{2}}.
\end{align}
By definition, $\Exs[f(\nu)] = \MYRAD(\YSET)$, so that setting $t =
\sqrt{2 (1 + \HACKKAP) \, \log(\numproj \numobs)}$ yields the bound
tail bound \mbox{$\mprob[\PLGOODONE^c] \leq \frac{1}{(\numproj
    \numobs)^\HACKKAP}\}$,} as claimed.\\

Next we control the probability of the event $\PLGOODTWO^c$.  The
function $g$ from equation~\eqref{EqnDefnG} is clearly convex in the
vector $\nu$; we now show that it is also Lipschitz with constant
$1/\sqrt{\numproj}$.  Indeed, for any two vectors $\nu, \nu' \in \{-1,
1\}^\usedim$, we have
\begin{align*}
|g(\nu) - g(\nu')| \leq \Exs_{\rade{}, P} \Big[ \sup_{y \in \YSET}
  \inprod{\frac{1}{\numproj} \sum_{i=1}^\numproj \rade{i} \diag (\nu -
    \nu') \sqrt{\numobs} H^T p_i}{z} \Big] \; & \leq \Exs_{\rade{}, P}
\| (\diag(\nu - \nu')) \, \sum_{i=1}^\numproj \rade{i} \sqrt{\numobs}
H^T p_i\|_2.
\end{align*}
Introducing the shorthand $\Delta = \diag(\nu - \nu')$ and
$\sketchtil_i = \sqrt{\numobs} H^T p_i$, Jensen's inequality yields
\begin{align*}
|g(\nu) - g(\nu')|^2 \; \leq \; \frac{1}{\numproj^2} \Exs_{\rade{}, P}
\| \Delta \, \sum_{i=1}^\numproj \rade{i} \sketchtil_i\|^2_2 & =
\frac{1}{\numproj^2} \trace \Big( \Delta \, \Exs_{P}
\big[\sum_{i=1}^\numproj \sketchtil_i \sketchtil_i^T \big] \Delta
\Big) \\
& = \frac{1}{\numproj} \trace \Big( \Delta^2 \; \diag (\Exs_{P} \big[
  \frac{1}{\numproj} \sum_{i=1}^\numproj \sketchtil_i \sketchtil_i^T
  \big] \big) \Big).
\end{align*}
By construction, we have $|\sketchtil_{ij}| = 1$ for all $(i,j)$,
whence $\diag (\Exs_{P} \big[ \frac{1}{\numproj} \sum_{i=1}^\numproj
  \sketchtil_i \sketchtil_i^T \big] \big) = I_{\numobs \times
  \numobs}$.  Since $\trace(\Delta^2) = \|\nu - \nu'\|_2^2$, we have
established that $|g(\nu) - g(\nu')|^2 \leq \frac{\|\nu -
  \nu'\|_2^2}{\numproj}$, showing that $g$ is a
$1/\sqrt{\numproj}$-Lipschitz function.  By standard concentration
results~\cite{Ledoux01}, we conclude that
\begin{align*}
\mprob[ \PLGOODTWO^c] \; = \; \mprob \big[ g(\nu) \geq \Exs[g(\nu)] +
  \frac{\delta}{\trunlev} \big] & \leq \CEXP{- \frac{\numproj
    \delta^2}{4 \trunlev^2}},
\end{align*}
as claimed.

\section{Proof of Lemma~\ref{LemmaInclusion}}
\label{AppSharpen}

By the inclusion~\eqref{EqnMertInclusions}(a), we have $\sup_{z \in
  \YSET_1} | z^T Q z | \leq \sup_{z \in \clconv (\YSET_0)} |z^T Q z|$.
Any vector \mbox{$v \in \conv(\YSET_0)$} can be written as a convex
combination of the form $v = \sum_{i=1}^T \alpha_i z_i$, where the
vectors $\{z_i\}_{i=1}^T$ belong to $\YSET_0$ and the non-negative
weights $\{\alpha_i\}_{i=1}^T$ sum to one, whence
\begin{align*}
|v^T Q v| & \leq \sum_{i=1}^T \sum_{j=1}^T \alpha_i \alpha_j \,
\big|z_i^T Q z_j \big| \\
& \leq \frac{1}{2} \max_{i,j \in [T]} \big|(z_i + z_j)^T Q (z_i + z_j)
- z_i^T Q z_i - z_j^T Q z_j \big| \\
& \leq \frac{3}{2} \sup_{z \in \SetDiff{\YSET_0}} |z^T Q z|.
\end{align*}
Since this upper bound applies to any vector $v \in \conv(\YSET_0)$, it
also applies to any vector in the closure, whence
\begin{align}
\label{EqnSelfBoundingFirst}
 \sup_{z \in \YSET_1} | z^T Q z | & \leq \; \sup_{z \in
   \clconv(\YSET_0)} |z^T Q z| \; \leq \; \frac{3}{2} \sup_{z \in
   \SetDiff{\YSET_0}} |z^T Q z|.
\end{align}

Now for some $\epsilon \in (0, 1]$ to be chosen, let $\{z^1,
\ldots,z^M \}$ be an $\epsilon$-covering of the set $\SetDiff{\YSET_0}$
in Euclidean norm.  Any vector $z \in \SetDiff{\YSET_0}$ can be written
as $z = z^j + \Delta$ for some $j \in [M]$, and some vector with
Euclidean norm at most $\epsilon$.  Moreover, the vector $\Delta \in
\SetDiffTwo{\YSET_0}$, whence
\begin{align}
\sup_{z \in \SetDiff{\YSET_0}} |z^T Q z| & \leq \max_{j \in [M]} |
(z^j)^T Q z^j | + 2 \sup_{ \substack{ \Delta \in \SetDiffTwo{\YSET_0}
    \\ \|\Delta\|_2\le \epsilon}} \; \max_{j \in [M]} |\Delta^T Q z^j|
+ \sup_{ \substack{ \Delta \in \SetDiffTwo{\YSET_0} \\ \|\Delta\|_2\le
    \epsilon}} |\Delta^T Q \Delta|.
\end{align}
Since $z^j \in \YSET_0 \subseteq \SetDiffTwo{\YSET_0}$, we have
\begin{align*}
\sup_{z \in \SetDiff{\YSET_0}} |z^T Q z| & \leq \max_{j \in [M]}
|(z^j)^T Q z_j| + 2 \sup_{ \substack{\Delta, \Delta' \in
    \SetDiffTwo{\YSET_0} \\ \|\Delta\|_2 \leq \epsilon}} |\Delta^T Q
\Delta' + \sup_{ \substack{ \Delta \in \SetDiffTwo{\YSET_0}
    \\ \|\Delta\|_2 \leq \epsilon}} |\Delta^T Q \Delta| \\
& \leq \max_{j \in [M]} |(z^j)^T Q z_j| + 3 \sup_{ \substack{\Delta,
    \Delta' \in \SetDiffTwo{\YSET_0} \\ \|\Delta\|_2 \leq \epsilon}}
|\Delta^T Q \Delta'| \\
& \leq \max_{j \in [M]} |(z^j)^T Q z_j| + 3 \epsilon \sup_{\substack{
    \Delta \in \SphereProj{\SetDiffTwo{\YSET_0}} \\ \Delta' \in
    \SetDiffTwo{\YSET_0}}} |\Delta^T Q \Delta'| \\
& \leq \max_{j \in [M]} |(z^j)^T Q z_j| + 3 \epsilon \sup_{\Delta,
  \Delta' \in \alpha \YSET_1} |\Delta^T Q \Delta'|,
\end{align*}
where the final inequality makes use of the
inclusions~\eqref{EqnMertInclusions}(b) and (c).  Finally, we observe
that
\begin{align*}
\sup_{\Delta,\Delta' \in \alpha \YSET_1} |\Delta^T Q \Delta'| & =
\sup_{\Delta,\Delta' \in \alpha \YSET_1} \frac{1}{2}
|(\Delta+\Delta')^T Q (\Delta+\Delta')^T - \Delta Q \Delta - \Delta' Q
\Delta' | \\
& \leq \frac{1}{2} \big \{ 4 + 1 + 1 \big \} \sup_{\Delta \in \alpha
  \YSET_1} |\Delta^T Q \Delta| \\
& = 3 \alpha^2 \sup_{z \in \YSET_1} |z^T Q z|,
\end{align*}
where we have used the fact that $\frac{\Delta + \Delta'}{2} \in
\alpha \YSET_1$, by convexity of the set $\alpha \YSET_1$.

Putting together the pieces, we have shown that
\begin{align*}
 \sup_{z \in \YSET_1} | z^T Q z | & \leq \; \frac{3}{2} \Big \{ \max_{j
   \in [M]} |(z^j)^T Q z_j| + 9 \epsilon \alpha^2 \sup_{\Delta \in
   \YSET_1} |\Delta^T Q \Delta| \Big \}.
\end{align*}
Setting $\epsilon=\frac{1}{27 \alpha^2}$ ensures that $9 \epsilon
\alpha^2 < 1/3$, and hence the claim~\eqref{EqnMertDiscretize} follows
after some simple algebra.

\section{A technical inclusion lemma}
\label{AppLemSandwich}

Recall the sets $\YSET_1(\rdim)$ and $\YSET_0(\rdim)$ previously
defined in equations~\eqref{EqnMertOne} and~\eqref{EqnMertZero}.
\blems
\label{LemSandwich}
We have the inclusion
\begin{align}
\label{EqnSandwich}
\YSET_1(\rdim) \; \subseteq \; \clconv \big( \YSET_0(\rdim) \big),
\end{align}
where $\clconv$ denotes the closed convex hull.
\elems
\HACKPROOF
Define the support functions $\phi_0(X) = \sup_{\Delta \in \YSET_0}
\tracer{X}{\Delta}$ and $\phi_1(X) = \sup_{\Delta \in \YSET_1}
\tracer{X}{\Delta}$.  It suffices to show that $\phi_1(X) \leq 3
\phi_0(X)$ for each $X \in \SYMMAT{\matdim}$.  The Frobenius norm,
nuclear norm and rank are all invariant to unitary transformation, so
we may take $X$ to be diagonal without loss of generality.  In this
case, we may restrict the optimization to diagonal matrices $\Delta$,
and note that
\begin{align*}
\frobnorm{\Delta} = \sqrt{\sum_{j=1}^\matdim \Delta^2_{jj}}, \quad
\mbox{and} \quad \nucnorm{\Delta} = \sum_{j = 1}^\usedim |\Delta_{jj}|.
\end{align*}
 Let $S$ be the indices of the $\radfloor$ diagonal elements
that are largest in absolute value.  It is easy to see that
\begin{align*}
\phi_0(X) & = \sqrt{\sum_{j \in S} X_{jj}^2}.
\end{align*}
On the other hand, for any index $k \notin S$, we have $|X_{kk}| \leq
|X_{jj}|$ for $j \in S$, and hence
\begin{align*}
\max_{k \notin S} |X_{kk}| \; \leq \; \frac{1}{\radfloor} \sum_{j \in
  S} |X_{jj}| \; \leq \frac{1}{\sqrt{\radfloor}} \sqrt{\sum_{j \in S}
  X^2_{jj}}
\end{align*}
Using this fact, we can write
\begin{align*}
\phi_1(X) & \leq \sup_{ \sum_{j \in S} \Delta_{jj}^2 \leq 1} \sum_{j
  \in S} \Delta_{jj} X_{jj} + \sup_{ \sum_{k \notin S} |\Delta_{kk}|
  \leq \sqrt{\rad}} \sum_{k \notin S} \Delta_{kk} X_{kk} \\
& = \sqrt{\sum_{j \in \Sset} X_{jj}^2} + \sqrt{\rad} \max_{k \notin S}
|X_{kk}| \\
 & \leq \big(1 + \frac{\sqrt{\rad}}{\sqrt{\radfloor}} \big)
\sqrt{\sum_{j \in \Sset} X_{jj}^2} \\
& \leq 3 \phi_0(X),
\end{align*}
as claimed.
\HACKENDPROOF
%

%



\bibliography{mert_super}


\end{document}